\documentclass[12pt]{article}
\usepackage{a4wide}
\usepackage{epsfig}
\newcommand{\oone}{\hbox{$1\kern-2.5pt\hbox{\rm l}$}}
\newcommand{\ssigma}{\hbox{$\kern2.5pt\vrule height4pt\kern-2.5pt\sigma$}}

\newcommand{\GeV}{{\rm\,GeV}}
\newcommand\pfrac[2]{\left(\frac{#1}{#2}\right)}
\newcommand{\Li}{{\rm Li}}
\newcommand{\imag}{\mathop{\rm Im}\nolimits}
\newcommand{\real}{\mathop{\rm Re}\nolimits}
\newcommand{\eps}{\varepsilon}
\newcommand{\cz}{\chi_Z}
\newcommand{\proj}{{\mit\Pi}}
\newcommand{\sxi}{\sqrt\xi}

\begin{document}
\strut\vspace{-3truecm}
\begin{flushright}
MZ-TH/02-29\\
hep-ph/0212039\\
Dec 2002
\end{flushright}

\begin{center}
{\Huge Polarization effects in\\[3pt]$e^+e^-$ annihilation processes}\\[12pt]
{\large Stefan Groote}\\[7pt]
Institut f\"ur Physik der Johannes-Gutenberg-Universit\"at,\\[3pt]
Staudinger Weg 7, 55099 Mainz, Germany
%\\[12pt]{\em Tartu \"Ulikool, 14ndal mail 2002}
\end{center}

\begin{abstract}
We present analytic results for first order radiative QCD corrections to
$e^+e^-$ annihilation processes into quarks and gluons under the special
aspect of polarized final state particles. This aspect involves single-spin
polarization, gluon polarization, and the correlation between quark spins
for massive final state quarks.
\end{abstract}

\tableofcontents

\vfill\noindent
{\em Invited talk given at the F\"u\"usika Instituut,\\
  Tartu \"Ulikool, Estonia, May 14th, 2002}
\newpage

\section{Introduction}
Quark mass effects play an important role in the production of quarks and
gluons in $e^+e^-$ annihilations. Jet definition schemes, event shape
variables, heavy flavour momentum correlations and the polarization of quarks
and gluons are affected by the presence of quark masses for charm and bottom
quarks even when they are produced at the scale of the $Z^0$
mass~\cite{BBU,Rodrigo,Nason}. A careful investigation of quark mass effects
in $e^+e^-$ annihilations may even lead to an alternative determination of the
quark mass values~\cite{BBU,Rodrigo,Nason,Grunberg}.

\vspace{7pt}
There is an obvious interest in quark mass effects for $t\bar t$ production
where quark mass effects cannot be neglected in the envisaged range of
energies to be covered by the Next Linear Collider (NLC). The longitudinal
polarization of massive quarks affects the shape of the energy spectrum of
their secondary decay leptons, the longitudinal spin-spin correlation effects
in pair produced quarks and antiquarks will lead to correlation effects of the
energy spectra of their secondary decay leptons and antileptons. Quark mass
effects are important in the $m\to 0$ calculation of radiative corrections to
quark polarization variables because residual mass effects change the naive
no-flip pattern of the $m=0$ polarization results~\cite{LeeNauSmi,Falk}.

\vspace{7pt}
In this talk I present $O(\alpha_s)$ radiative corrections to different
polarization observables, namely the longitudinal and transverse single-spin
polarization of the quark~\cite{KPT94,GKT96,GK96,GKT97}, the polarization of
the gluon~\cite{GKL97,GKL99}, and the longitudinal spin-spin
correlation~\cite{GKL98,GKL00} for massive quark pairs produced in $e^+e^-$
annihilations, always including the polar angle dependence. As a byproduct of
our calculations I discuss the $m\to 0$ limit and the role of the
$O(\alpha_s)$ residual mass effects. I explain how residual mass effects
contribute to the various spin-flip and no-flip terms in the $m\to 0$ limit
for each of the three structure functions that describe the polar angle
dependence.

\section{The joint quark-antiquark density matrix}
Let us start with the joint quark-antiquark density matrix
$d\ssigma^\alpha=d\sigma^\alpha_{\lambda_1\lambda_2;\lambda'_1\lambda'_2}$
where $\lambda_1$ and $\lambda_2$ denote the helicities of the quark and
antiquark, respectively. In the applications covered by this talk, only the
diagonal case $\lambda_1=\lambda'_1$ and $\lambda_2=\lambda'_2$ is of
importance. The label $\alpha$ specifies the polarization of the initial
$\gamma^*$, $Z$ or interference contributions thereof. This index will be
specified later on by the labels $U$ (unpolarized transverse), $L$
(longitudinal), $F$ (forward--backward), and the two labels $A$ and $I$ for
the case of transverse polarization of the final state quarks.

\vspace{7pt}
The diagonal part of the differential joint density matrix can be represented
in terms of its components along the products of the unit matrix and the
$z$-components of the Pauli matrix $\ssigma_3$
($\ssigma_3=\hat p_1\vec{\ssigma}$ for the quark and
$\ssigma_3=\hat p_2\vec{\ssigma}$ for the antiquark, 
$\hat p_i=\vec p_i/|\vec p_i|$). One has
\begin{equation}
d\ssigma^\alpha=\frac14\left(d\sigma_\alpha\oone\otimes\oone
  +d\sigma_\alpha^{(\ell_1)}\ssigma_3\otimes\oone
  +d\sigma_\alpha^{(\ell_2)}\oone\otimes\ssigma_3
  +d\sigma_\alpha^{(\ell_1\ell_2)}\ssigma_3\otimes\ssigma_3\right)
\end{equation}
as formulated for longitudinal polarization, the first and the second Pauli
matrices stand for the quark and the antiquark, respectively. An alternative
but equivalent representation of the longitudinal spin contributions can be
written down in terms of the longitudinal spin components
$s_1^\ell=2\lambda_1$ and $s_2^\ell=2\lambda_2$ with $s_1^\ell,s_2^\ell=\pm 1$
(or $s_1^\ell,s_2^\ell\in\{\uparrow,\downarrow\}$). The spin dependent parts
of the differential cross section are given by
\begin{equation}\label{eqnkd1}
d\sigma_\alpha(s_1^\ell,s_2^\ell)=\frac14\left(d\sigma_\alpha
+d\sigma^{(\ell_1)}_\alpha s_1^\ell+d\sigma^{(\ell_2)}_\alpha s_2^\ell
+d\sigma^{(\ell_1\ell_2)}_\alpha s_1^\ell s_2^\ell\right).
\end{equation}
Eq.~(\ref{eqnkd1}) is easily inverted,
\begin{eqnarray}
d\sigma_\alpha&=&d\sigma_\alpha(\uparrow\uparrow)
  +d\sigma_\alpha(\uparrow\downarrow)+d\sigma_\alpha(\downarrow\uparrow)
  +d\sigma_\alpha(\downarrow\downarrow),\nonumber\\
d\sigma_\alpha^{(\ell_1)}&=&d\sigma_\alpha(\uparrow\uparrow)
  +d\sigma_\alpha(\uparrow\downarrow)-d\sigma_\alpha(\downarrow\uparrow)
  -d\sigma_\alpha(\downarrow\downarrow),\nonumber\\
d\sigma_\alpha^{(\ell_2)}&=&d\sigma_\alpha(\uparrow\uparrow)
  -d\sigma_\alpha(\uparrow\downarrow)+d\sigma_\alpha(\downarrow\uparrow)
  -d\sigma_\alpha(\downarrow\downarrow),\nonumber\\
d\sigma_\alpha^{(\ell_1\ell_2)}&=&d\sigma_\alpha(\uparrow\uparrow)
  -d\sigma_\alpha(\uparrow\downarrow)-d\sigma_\alpha(\downarrow\uparrow)
  +d\sigma_\alpha(\downarrow\downarrow).
\end{eqnarray}
The first line is the unpolarized contribution, the second line the
contribution due to a single polarized quark, while the last line is the
correlation between quark and antiquark. While all these contributions will
be considered in the following, the polarisation of a single antiquark
(third line) is not considered because it is equivalent to the case of a
single polarized quark.

\subsection{The differential cross section}
According to Fermi's ``golden rule'' the differential cross section is given
by
\begin{equation}
d\sigma=2\pi|T_{fi}|^2dPS.
\end{equation}
Using conventional Feynman rules, especially
\begin{equation}
-ieQ_f\gamma_\mu\quad\mbox{and}\quad
-ie(v_f\gamma_\mu-a_f\gamma_\mu\gamma_5)
\end{equation}
for the vertex of the fermion line with a photon or a $Z$ boson, respectively,
and
\begin{equation}
\frac{-ig^{\mu\nu}}{q^2}\quad\mbox{and}\quad
\chi_Z(q)\frac{-ig^{\mu\nu}}{q^2}
\end{equation}
for the boson propagators with momentum $q$, one can write
\begin{eqnarray}
\lefteqn{T_{fi}\ =\ \bar v(p_2)\Big[-ieQ_f\gamma^\mu\Big]u(p_1)
  \frac{-i}{q^2}\bar v(p_+)\Big[-ieQ_e\gamma_\mu\Big]u(p_-)\,+}\nonumber\\&&
  +\bar v(p_2)\Big[-ie(v_f\gamma^\mu-a_f\gamma^\mu\gamma_5)\Big]u(p_1)
  \chi_Z(q)\frac{-i}{q^2}
  \bar v(p_+)\Big[-ie(v_e\gamma_\mu-a_e\gamma_\mu\gamma_5)\Big]u(p_-)\qquad
\end{eqnarray}
where $p_-$ and $p_-$ are the moments of the electron and positron, $p_1$ and
$p_2$ are the moments of the quark and antiquark, and $q=p_-+p_+=p_1+p_2$ is
the moment of the boson. Possible polarisation degrees of freedom enter the
squared matrix element by replacing for instance $u(p_1)\bar u(p_1)$ by
$(\gamma_\mu p_1^\mu+m)(1+\gamma_5\gamma_\nu s_1^\nu)/2$ instead of summing
over the polarisations of the final state. $m$ is the mass of the quark (and
antiquark). The squared matrix element reads
\begin{equation}\label{Tfi2}
|T_{fi}|^2=g_{ij}L_{\mu\nu}^iH^{j\mu\nu}
\end{equation}
where $L_{\mu\nu}^i$ contains all elements due to electron and positron
(lepton tensor), $H^{j\mu\nu}$ contains all elements due to the quark and
antiquark (hadron tensor), and $g_{ij}$ contains the elements due to the
intemediate bosons (photon and $Z$ boson) which mix up. These parts will be
considered in turn.

\subsection{The hadron tensor}
The hadron tensor is determined by the hadron dynamics, i.e.\ by the
current-induced production of a quark-antiquark pair which, in the
$O(\alpha_s)$ case, is followed by gluon emission. In the $O(\alpha_s)$ case
one also has to add the one-loop contribution. The index $i=1,2,3,4$ in
Eq.~(\ref{Tfi2}) specifies the current composition in terms of the parity-even
(for $i=1,2$) and parity-odd ($i=3,4$) products of the vector and the axial
vector currents according to
\begin{eqnarray}\label{H1234}
H^1_{\mu\nu}=\frac12(H^{VV}_{\mu\nu}+H^{AA}_{\mu\nu}),&&
H^2_{\mu\nu}=\frac12(H^{VV}_{\mu\nu}-H^{AA}_{\mu\nu}),\nonumber\\
H^3_{\mu\nu}=\frac{i}2(H^{VA}_{\mu\nu}-H^{A\,V}_{\mu\nu}),&&
H^4_{\mu\nu}=\frac12(H^{VA}_{\mu\nu}+H^{A\,V}_{\mu\nu}).
\end{eqnarray}
In order to be specific, the unpolarized hadron tensor $H_{\mu\nu}^{VA}$ is
e.g.\ given by the matrix element
\begin{equation}
H_{\mu\nu}^{VA}=\langle q\bar q|J_\mu^V|0\rangle
  \langle 0|(J_\nu^A)^\dagger|q\bar q\rangle.
\end{equation}
In the two-body case $e^+e^-\to q\bar q$ the components of the cross section
are given by
\begin{equation}
\sigma_\alpha^{i(P)}=\frac{\pi\alpha^2v}{3q^4}
  H_\alpha^{i(P)}\qquad
  \left(\mbox{with\ }v=\sqrt{1-4m^2/q^2}\right)
\end{equation}
where $(P)$ stands for unpolarized, single-spin or spin-spin correlation
contributions. In the three-body case $e^+e^-\to q\bar qg$ the hadron tensor
components $H_\alpha^{i(P)}(y,z)$ are related to the components of the
differential cross section by
\begin{equation}\label{3to2}
\frac{d\sigma_\alpha^{i(P)}}{dy\,dz}=\frac{\pi\alpha^2v}{3q^4}
  \left\{\frac{q^2}{16\pi^2v}H_\alpha^{i(P)}(y,z)\right\}.
\end{equation}
The two energy-type variables $y=1-2p_1q/q^2$ and $z=1-2p_2q/q^2$ are used
as kinematic variables. Note that the three-body helicity structure functions
$H_\alpha^{i(P)}(y,z)$ have a different dimension than their two-body
counterparts in Eq.~(\ref{3to2}) which we indicate by explicitly referring to
the $(y,z)$-dependence of the three-body structure functions. An example for
the phase space in terms of $y$ and $z$ is shown in Fig.~\ref{fig4}.

\subsection{The electro-weak form factors}
The second building block $g_{ij}$ ($i,j=1,2,3,4$) specifies the electro-weak
model dependence of the $e^+e^-$ cross section. For the present discussion we
need the components $g_{11}$, $g_{12}$, $g_{43}$ and $g_{44}$. They are given
by
\begin{eqnarray}
g_{11}&=&Q_f^2-2Q_fv_ev_f\real\cz+(v_e^2+a_e^2)(v_f^2+a_f^2)|\cz|^2,\nonumber\\
g_{12}&=&Q_f^2-2Q_fv_ev_f\real\cz+(v_e^2+a_e^2)(v_f^2-a_f^2)|\cz|^2,\nonumber\\
g_{13}&=&-2Q_fv_ea_f\imag\cz,\nonumber\\
g_{14}&=&2Q_fv_ea_f\real\cz-2(v_e^2+a_e^2)v_fa_f|\cz|^2,\\[12pt]
g_{21}&=&q_f^2-2Q_fv_ev_f\real\cz+(v_e^2-a_e^2)(v_f^2+a_f^2)|\cz|^2,\nonumber\\
g_{22}&=&q_f^2-2Q_fv_ev_f\real\cz+(v_e^2-a_e^2)(v_f^2-a_f^2)|\cz|^2,\nonumber\\
g_{23}&=&-2Q_fv_ea_f\imag\cz,\nonumber\\
g_{24}&=&2Q_fv_ea_f\real\cz-2(v_e^2-a_e^2)v_fa_f|\cz|^2,\\[12pt]
g_{31}&=&-2Q_fa_ev_f\imag\cz,\nonumber\\
g_{32}&=&-2Q_fa_ev_f\imag\cz,\nonumber\\
g_{33}&=&2Q_fa_ea_f\real\cz,\nonumber\\
g_{34}&=&2Q_fa_ea_f\imag\cz,\\[12pt]
g_{41}&=&2Q_fa_ev_f\real\cz-2v_ea_e(v_f^2+a_f^2)|\cz|^2,\nonumber\\
g_{42}&=&2Q_fa_ev_f\real\cz-2v_ea_e(v_f^2-a_f^2)|\cz|^2,\nonumber\\
g_{43}&=&2Q_fa_ea_f\imag\cz,\nonumber\\
g_{44}&=&-2Q_fa_ea_f\real\cz+4v_ea_ev_fa_f|\cz|^2
\end{eqnarray}
where $\cz(q^2)=gM_Z^2q^2/(q^2-M_Z^2+iM_Z\Gamma_Z)$, with $M_Z$ and $\Gamma_Z$
the mass and width of the $Z^0$ ($M_Z=91.1887\GeV$ and
$\Gamma_Z=2.487\GeV$~\cite{PDG}) and
$g=G_F(8\sqrt 2\pi\alpha)^{-1}\approx 4.49\times 10^{-5}\mbox{\rm GeV}^{-2}$.
$Q_f$ are the charges of the final state quarks to which the electro-weak
currents directly couple; $v_e$ and $a_e$, $v_f$ and $a_f$ are the
electro-weak vector and axial vector coupling constants. For example, in the
Weinberg--Salam model, one has $v_e=-1+4\sin^2\theta_W$, $a_e=-1$ for leptons,
$v_f=1-\frac83\sin^2\theta_W$, $a_f=1$ for up-type quarks ($Q_f=\frac23$), and
$v_f=-1+\frac43\sin^2\theta_W$, $a_f=-1$ for down-type quarks ($Q_f=-\frac13$).
The left- and right-handed coupling constants are then given by $g_L=v+a$ and
$g_R=v-a$, respectively. In the purely electromagnetic case one has
$g_{11}=g_{12}=g_{21}=g_{22}=Q_f^2$ and all other $g_{r'r}=0$. The terms
linear in $\real\cz$ and $\imag\cz$ come from $\gamma-Z^0$ interference,
whereas the terms proportional to $|\cz|^2$ originate from $Z$-exchange.

\newpage
\vspace{7pt}
The generalization of the case where one starts with longitudinally polarized
beams is straightforward and amounts to the replacement
\begin{equation}
g_{1i}\rightarrow(1-h^-h^+)g_{1i}+(h^--h^+)g_{4i}\qquad
g_{4i}\rightarrow(1-h^-h^+)g_{4i}+(h^--h^+)g_{1i}
\end{equation}
where $h^-$ and $h^+$ ($-1\le h^\pm\le+1$) denote the longitudinal
polarization of the electron and the positron beam, respectively. Clearly
there is no interaction between the beams when $h^+=h^-=\pm1$.

\subsection{The polar angle dependence}
As mentioned in the introduction, the polar angle dependence will be
determined. The rest frame of the boson is a natural choice for the
considerations. However, two coordinate systems are in use here, the {\em
event frame\/} for the outgoing particles and the {\em beam frame\/} for the
incoming particles. Transforming from one to the other system brings in a
relative polar angle $\theta$ (and in case of transverse spin vectors also a
azimuthal angle $\varphi$ which is averaged out normally). The polar angle
$\theta$ is given by the angle between the colliding electron beam direction
and the outgoing quark direction. Transformed to the event frame, the lepton
tensor $L_{\mu\nu}^i$ leads to the different angular distributions. While for
massless leptons the components $L_{\mu\nu}^2$ and $L_{\mu\nu}^3$ vanish,
one can decompose the remaining lepton tensor components according to
\begin{eqnarray}\label{decomp}
L^1&=&\frac{q^2}2\left\{\frac12(1+\cos^2\theta)\proj_U+\sin^2\theta\,\proj_L
  +\sin\theta\cos\theta\,\proj_I\right\},\nonumber\\
L^4&=&\frac{q^2}2\left\{\cos\theta\,\proj_F+\sin\theta\,\proj_A\right\}.
\end{eqnarray}
Remark that $\proj_I$ and $\proj_A$ depend linearly on $\sin\varphi$ and
$\cos\varphi$. The matrices $\proj_U$, $\proj_L$, $\proj_I$, $\proj_F$ and
$\proj_A$ are called {\it projectors\/} because in contracting the lepton
tensor with the hadron tensor they project out the contribution of the hadron
tensor to the different angle dependences. Explicit projectors are
\begin{eqnarray}\label{ULFproj}
\proj_U^{\mu\nu}&=&-\hat g^{\mu\nu}-\frac{\hat p_1^\mu\hat p_1^\nu}{p_{1z}^2},
  \qquad
\proj_L^{\mu\nu}\ =\ \frac{\hat p_1^\mu\hat p_1^\nu}{p_{1z}^2},\qquad
\proj_F^{\mu\nu}\ =\ i\varepsilon^{\mu\nu\alpha\beta}
  \frac{\hat p_{1\alpha}q_\beta}{p_{1z}\sqrt{q^2}},\\
\proj_I^{\mu\nu}&=&s_1^\mu\hat p_1^\nu+\hat p_1^\mu s_1^\nu,\qquad
\proj_A^{\mu\nu}\ =\ i\eps^{\mu\nu\alpha\beta}q_\alpha s_{1\beta}\qquad
(\hat g^{\mu\nu}=g^{\mu\nu}-q^\mu q^\nu/q^2,\ \hat p_i=\vec p_i/|\vec p_i|)
\nonumber
\end{eqnarray}
The decomposition in Eq.~(\ref{decomp}) covers all possible angle dependences
which occur in the process with a single polarized quark. This decomposition
gives rise to the decomposition of the differential cross section according to
\begin{eqnarray}\label{angdecomp}
\frac{d\sigma^{(P)}}{d\cos\theta}
  &=&\frac38(1+\cos^2\theta)\sigma_U^{(P)}
  +\frac34\sin^2\theta\sigma_L^{(P)}
  +\frac34\cos\theta\sigma_F^{(P)}\,+\nonumber\\&&\qquad
  +\frac34\sin\theta\cos\theta\sigma_I^{(P)}
  +\frac34\sin\theta\sigma_A^{(P)}
\end{eqnarray}

\section{Unpolarized and polarized structure functions}
In Ref.~\cite{KPT94} the mean longitudinal polarization of the quark (i.e.\
the average over the polar angle $\theta$) was calculated, while in
Refs.~\cite{GKT96,GKT97} the polar angle dependence was determined, including
beam polarisation effects. The polar angle dependence of the transverse
polarisation was considered in Ref.~\cite{GK96}. In this section I present the
results and show a few instructive steps of the calculations.

\begin{figure}[ht]\begin{center}
\epsfig{figure=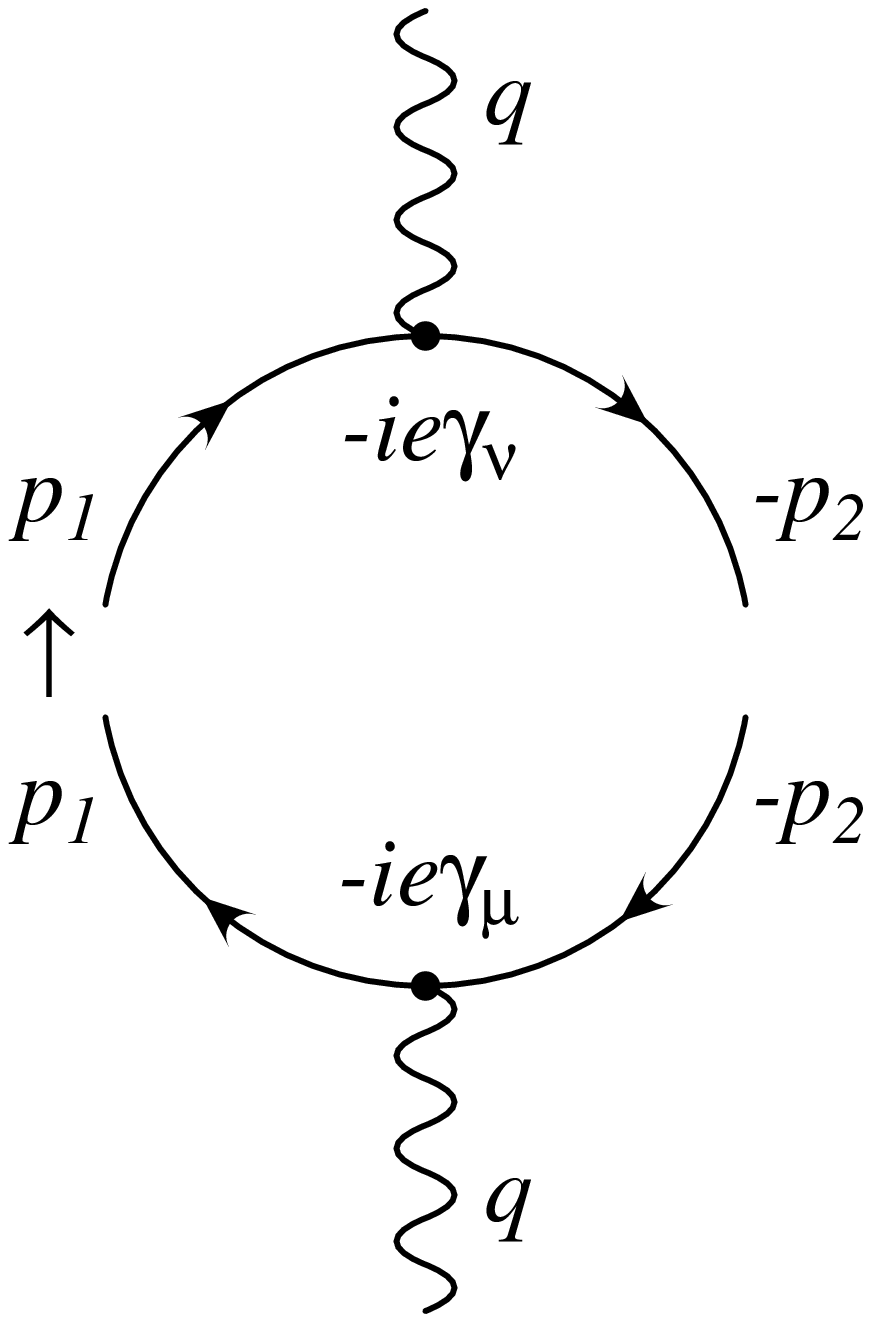, scale=0.3}\qquad
\epsfig{figure=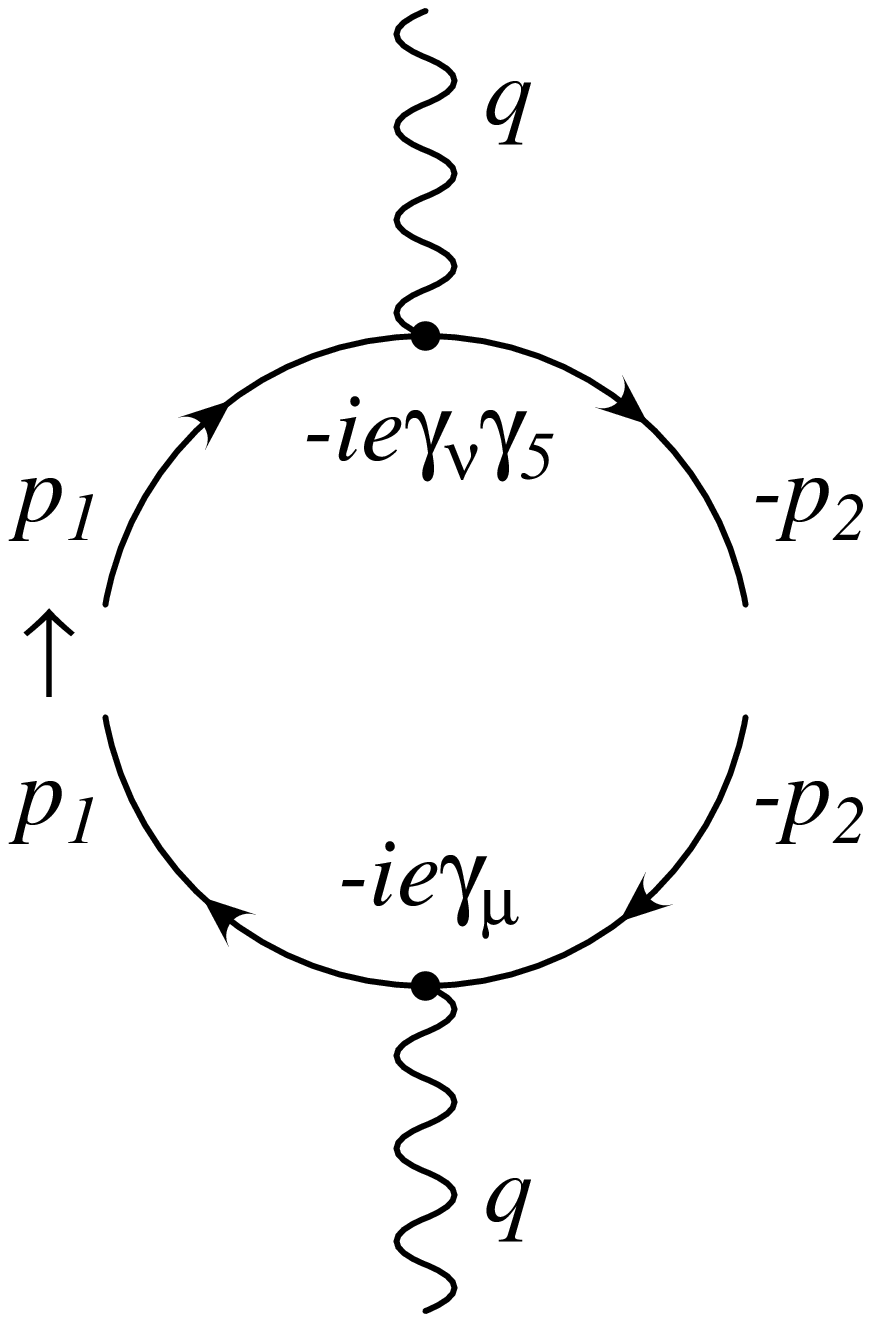, scale=0.3}\qquad
\epsfig{figure=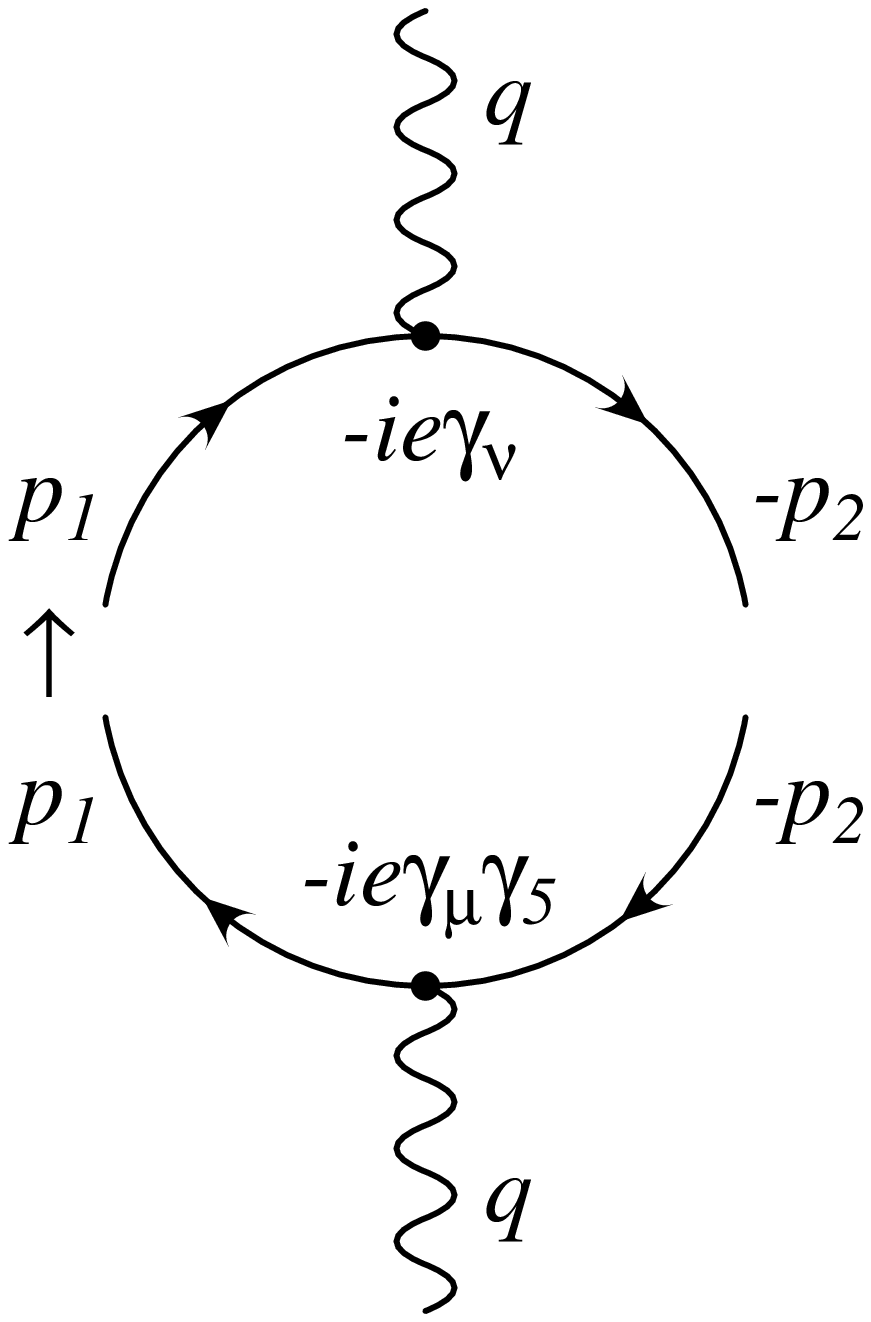, scale=0.3}\qquad
\epsfig{figure=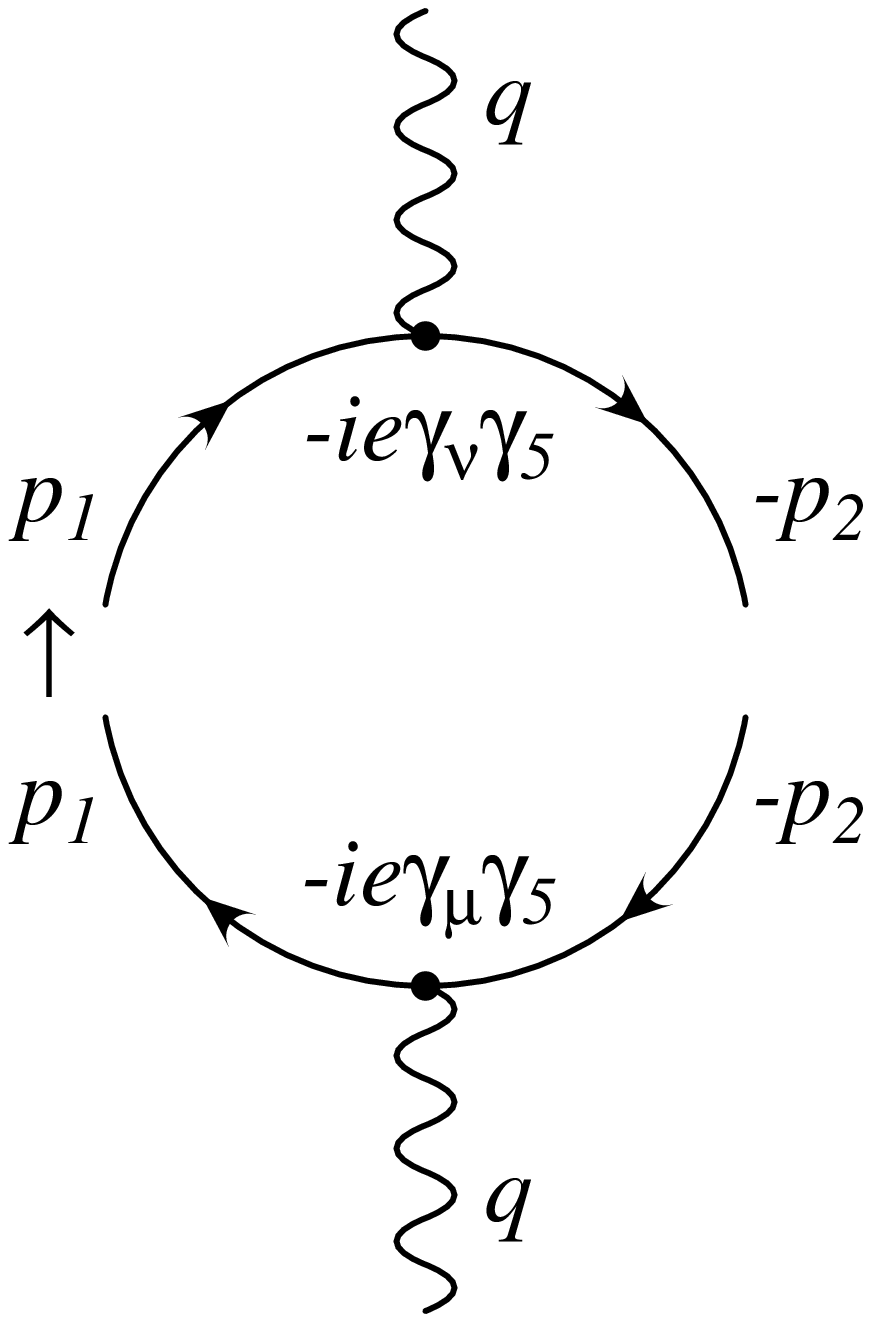, scale=0.3}
\caption{\label{fig1}Born term contributions to the single spin case}
\end{center}\end{figure}

\subsection{Born term contributions}
The Born term contributions are easy to calculate. The Feynman diagrams for
the hadronic part (in case of the single polarized case) are shown in
Fig.~\ref{fig1}. The unpolarized contributions are given by
\begin{eqnarray}
H_U^1({\it born\/})=2N_cq^2(1+v^2),&&
H_L^1({\it born\/})=N_cq^2(1-v^2)\ =\ H_L^2({\it born\/}),\nonumber\\[3pt]
H_U^2({\it born\/})=2N_cq^2(1-v^2),&&
H_F^4({\it born\/})=4N_cq^2v.
\end{eqnarray}
where $N_c$ is the number of colours. The longitudinally polarized
contributions read
\begin{eqnarray}
H_U^{4\ell}({\it born\/})=4N_cq^2v,&&
H_F^{1\ell}({\it born\/})=2N_cq^2(1+v^2),\nonumber\\[3pt]
H_L^{4\ell}({\it born\/})=0,&&
H_F^{2\ell}({\it born\/})=2N_cq^2(1-v^2).
\end{eqnarray}
For the transverse polarization we finally obtain
\begin{eqnarray}
H_I^{4\perp}({\it born\/})=-2N_cq^2v\sxi,&&
H_A^{1\perp}({\it born\/})=-2N_cq^2\sxi=H_A^{2\perp}({\it born\/}),\nonumber\\
&&H_A^{3N}({\it born\/})=2N_cq^2\sxi
\end{eqnarray}
where $\xi=4m^2/q^2=1-v^2$ and ``$\perp$'' and ``$N$'' indicate the transverse
polarisations perpendicular and normal to the event plane.

\begin{figure}[ht]\begin{center}
\epsfig{figure=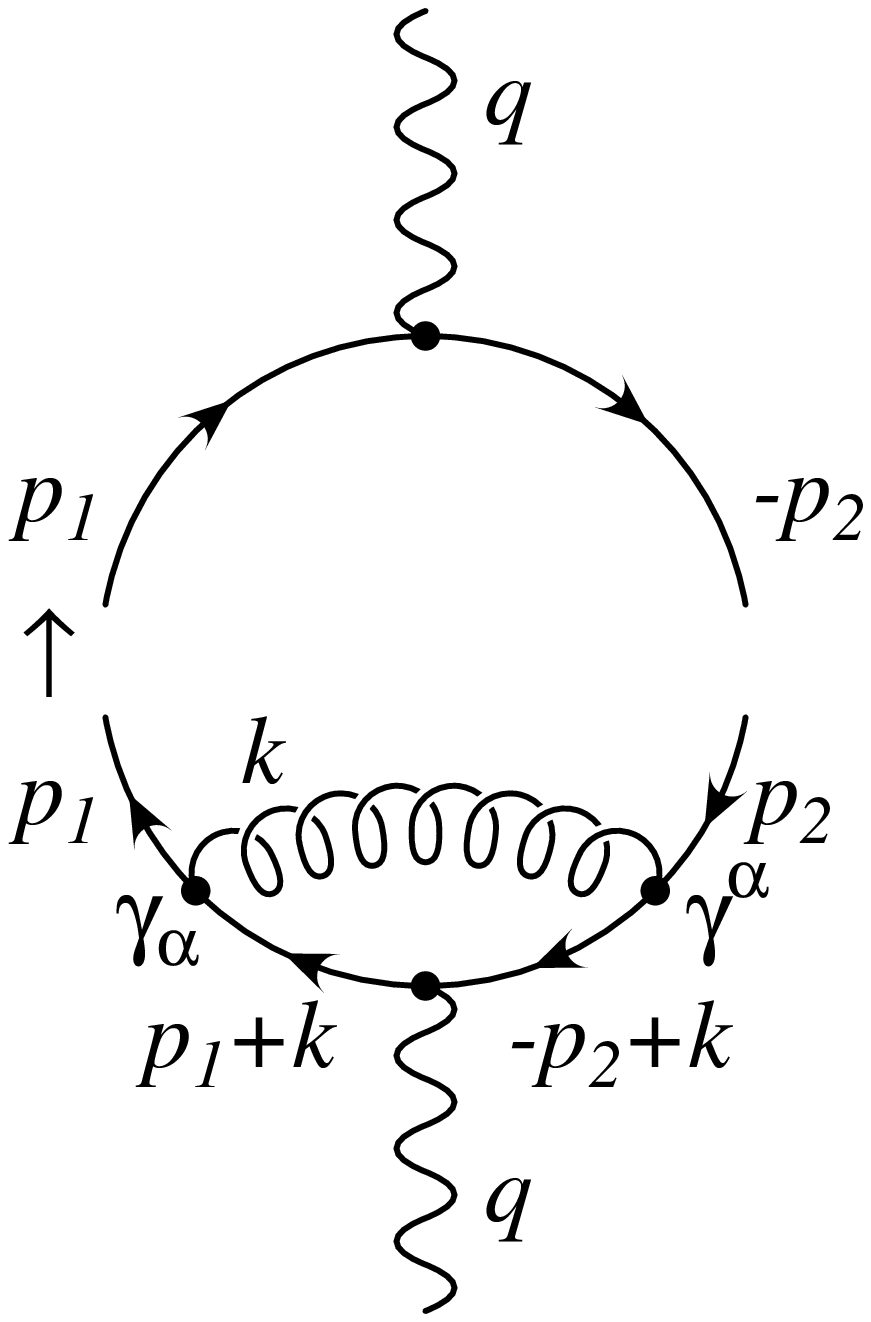, scale=0.3}\qquad
\epsfig{figure=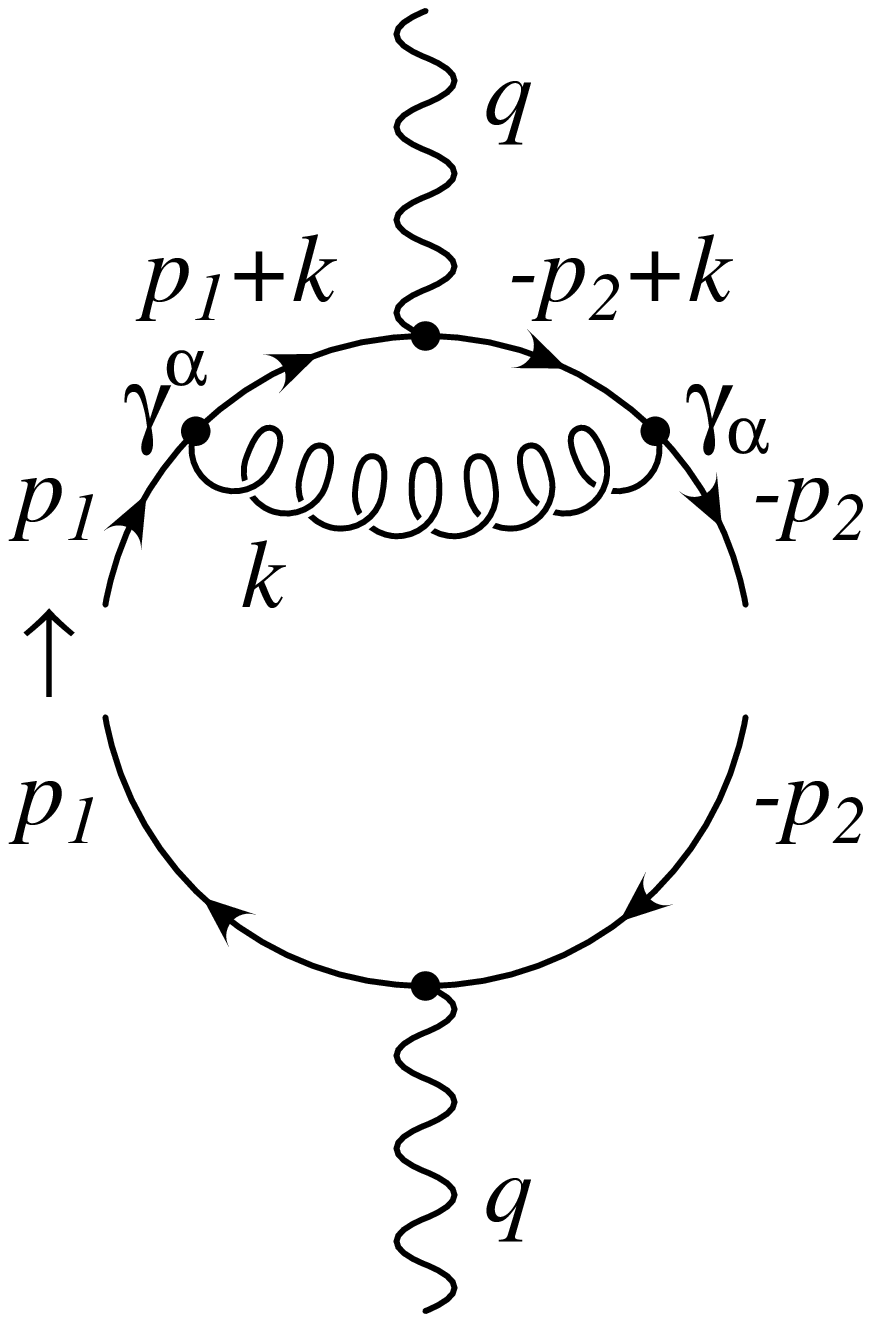, scale=0.3}
\caption{\label{fig2}Loop contributions to the single spin case}
\end{center}\end{figure}

\subsection{First order loop contributions}
The calculation of loop contributions can be formally be done by using the
vertex form factors, i.e.\ by replacing the vector and axial vector vertices
according to
\begin{equation}
\gamma^\mu\rightarrow(1+A)\gamma^\mu+B(p_1-p_2)^\mu,\qquad
\gamma^\mu\gamma_5\rightarrow
  \left((1+C)\gamma^\mu+D(p_1+p_2)^\mu\right)\gamma_5.
\end{equation}
The squared matrix element including first order loop corrections contains
the Born term contributions as well as first and second order contributions.
Only the first order contributions, as shown in Fig.~\ref{fig2}, are used
for a first order calculation. The non-vanishing contributions read
\begin{eqnarray}
H_U^1({\it loop\/})&=&4N_cq^2\left((1+v^2)\real A-2v^2\real B\right),
  \nonumber\\
H_U^2({\it loop\/})&=&4N_cq^2\left((1-v^2)\real A+2v^2\real B\right),
  \nonumber\\
H_L^1({\it loop\/})&=&2N_cq^2\left((1-v^2)\real A+v^2\real B\right)
  \ =\ H_L^2({\it loop\/}),\nonumber\\
H_F^4({\it loop\/})&=&8N_cq^2v\left(\real A-\real B\right),\nonumber\\[7pt]
H_U^{4\ell}({\it loop\/})&=&8N_cq^2v\left(\real A-\real B\right),\qquad
H_L^{4\ell}({\it loop\/})\ =\ 0,\nonumber\\
H_F^{1\ell}({\it loop\/})&=&4N_cq^2\left((1+v^2)\real A-2v^2\real B\right),
  \nonumber\\
H_F^{2\ell}({\it loop\/})&=&4N_cq^2\left((1-v^2)\real A+2v^2\real B\right),
  \nonumber\\[7pt]
H_I^{3\perp}({\it loop\/})&=&-2N_cq^2v\sxi(1+\xi)\imag B/\xi,\nonumber\\
H_I^{4\perp}({\it loop\/})&=&-2N_cq^2v\sxi
  \left(2\real A+(1-3\xi)\real B/\xi\right),\nonumber\\
H_A^{1\perp}({\it loop\/})&=&-2N_cq^2\sxi\left(\real A-v^2\real B/\xi\right)
  \ =\ H_A^{2\perp}({\it loop\/}),\nonumber\\[7pt]
H_I^{1N}({\it loop\/})&=&-2N_cq^2v\sxi(1-\xi)\imag B/\xi
  \ =\ H_I^{2N}({\it loop\/}),\nonumber\\
H_A^{3N}({\it loop\/})&=&2N_cq^2v\sxi
  \left(2\real A+(1-3\xi)\real B/\xi\right),\nonumber\\
H_A^{4N}({\it loop\/})&=&2N_cq^2v\sxi(1+\xi)\imag B/\xi,
\end{eqnarray}
the vertex form factors $A$ and $B$ are given by
\begin{eqnarray}
\real A&=&-\frac{\alpha_sC_F}{4\pi}
  \Bigg\{\left(2+\frac{1+v^2}v\ln\pfrac{1-v}{1+v}\right)
  \ln\pfrac{\Lambda q^2}{m^2}+3v\ln\pfrac{1-v}{1+v}+4\nonumber\\&&\qquad
  +\frac{1+v^2}v\left(\Li_2\pfrac{2v}{1+v}+\frac14\ln^2\pfrac{1-v}{1+v}
  -\frac{\pi^2}2\right)\Bigg\},\nonumber\\
\real B&=&\frac{\alpha_sC_F}{4\pi}\ \frac{1-v^2}v\ln\pfrac{1-v}{1+v},\qquad
\imag B\ =\ \frac{\alpha_sC_F}{4\pi}\ \frac{1-v^2}v\pi
\end{eqnarray}
were we already have replaced the axial-vector form factor $C$ by $A$ according
to $C=A-2B$, and where $m_G^2=\Lambda q^2$ is the squared mass of the gluon
which we use for the regularization of the first order tree contributions and
which is replaces the parameter $\eps=(4-D)/2$ of the dimensional
regularization by means of
$1/\eps-\gamma_E+\ln(4\pi\mu^2/m^2)\rightarrow\ln(\Lambda q^2/m^2)$.

\begin{figure}[ht]\begin{center}
\epsfig{figure=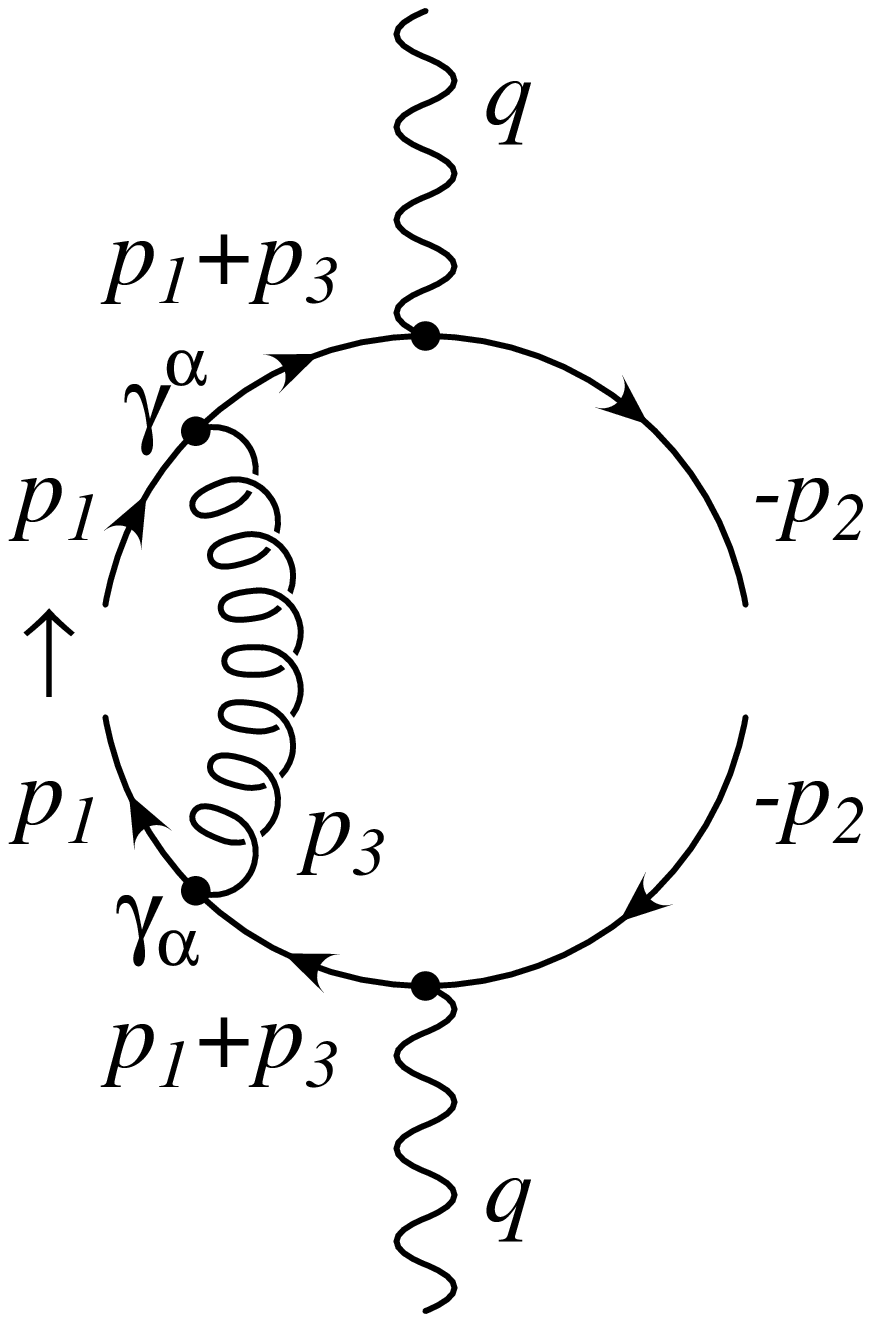, scale=0.3}\qquad
\epsfig{figure=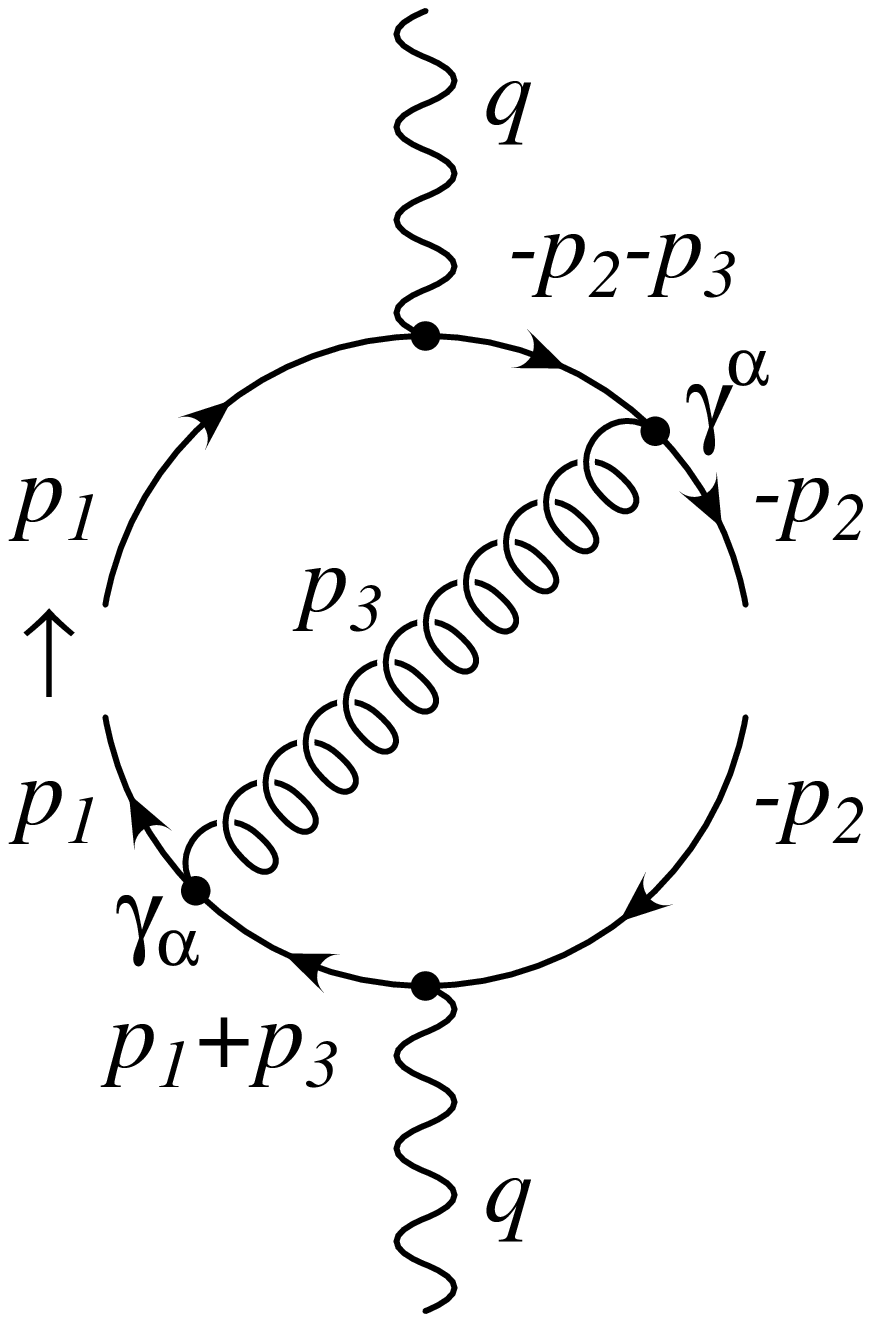, scale=0.3}\qquad
\epsfig{figure=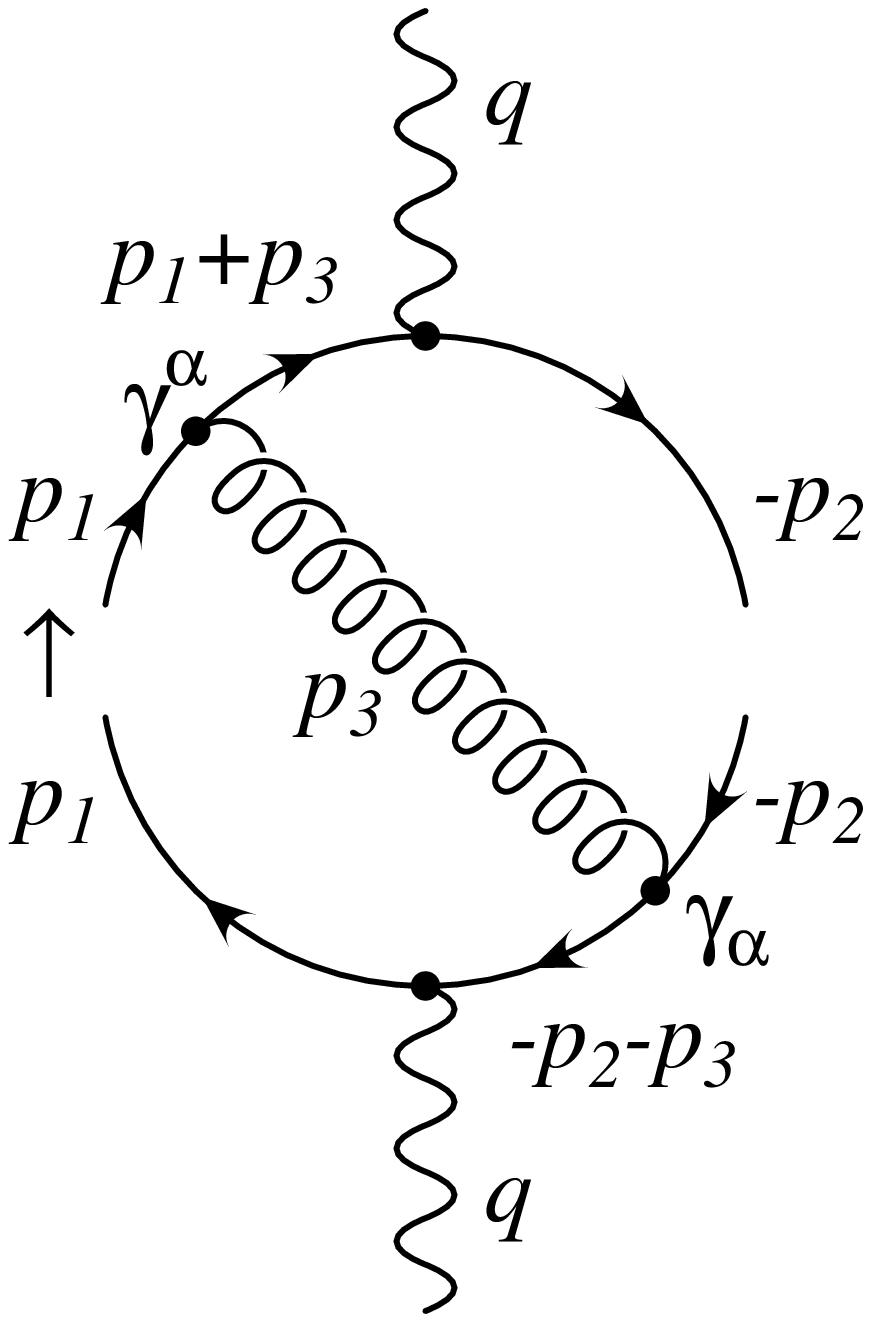, scale=0.3}\qquad
\epsfig{figure=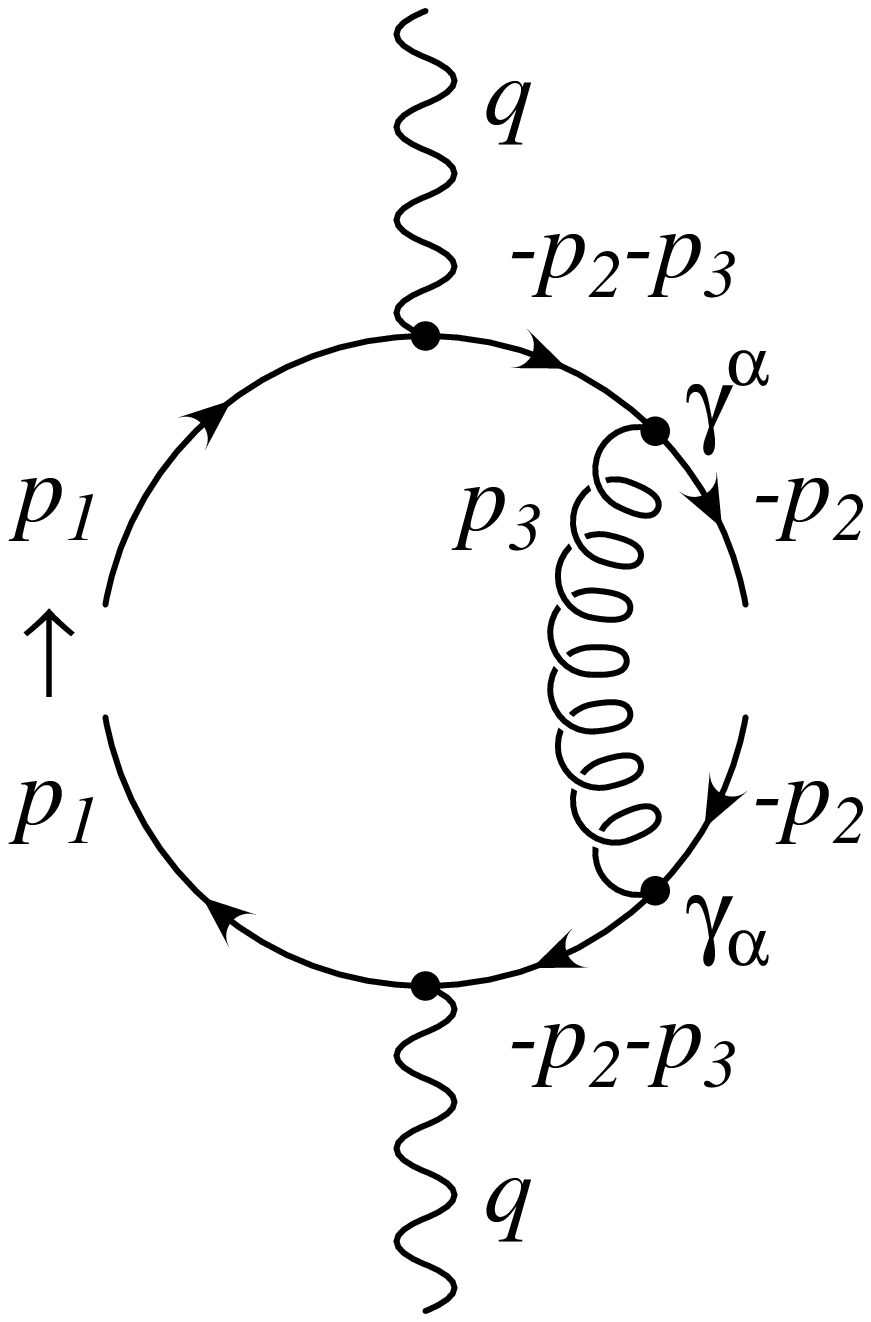, scale=0.3}
\caption{\label{fig3}tree graph contributions to the single spin case}
\end{center}\end{figure}

\subsection{First order tree graph contributions}
Even though ending up with a different final state, namely a three particle
state including quark, antiquark and gluon, the first order tree graph
contributions are needed for a full first order result. For vanishing gluon
momentum the (soft) gluon cannot be resolved by the detector, the event is
counted as two particle final state. In this soft region the infrared
singularity of the three graph contribution cancels the infrared singularity
of the loop contribution. This is a general result formulated in the
Lee--Nauenberg theorem~\cite{LeeNauSmi}.

\begin{figure}[t]\begin{center}
\epsfig{figure=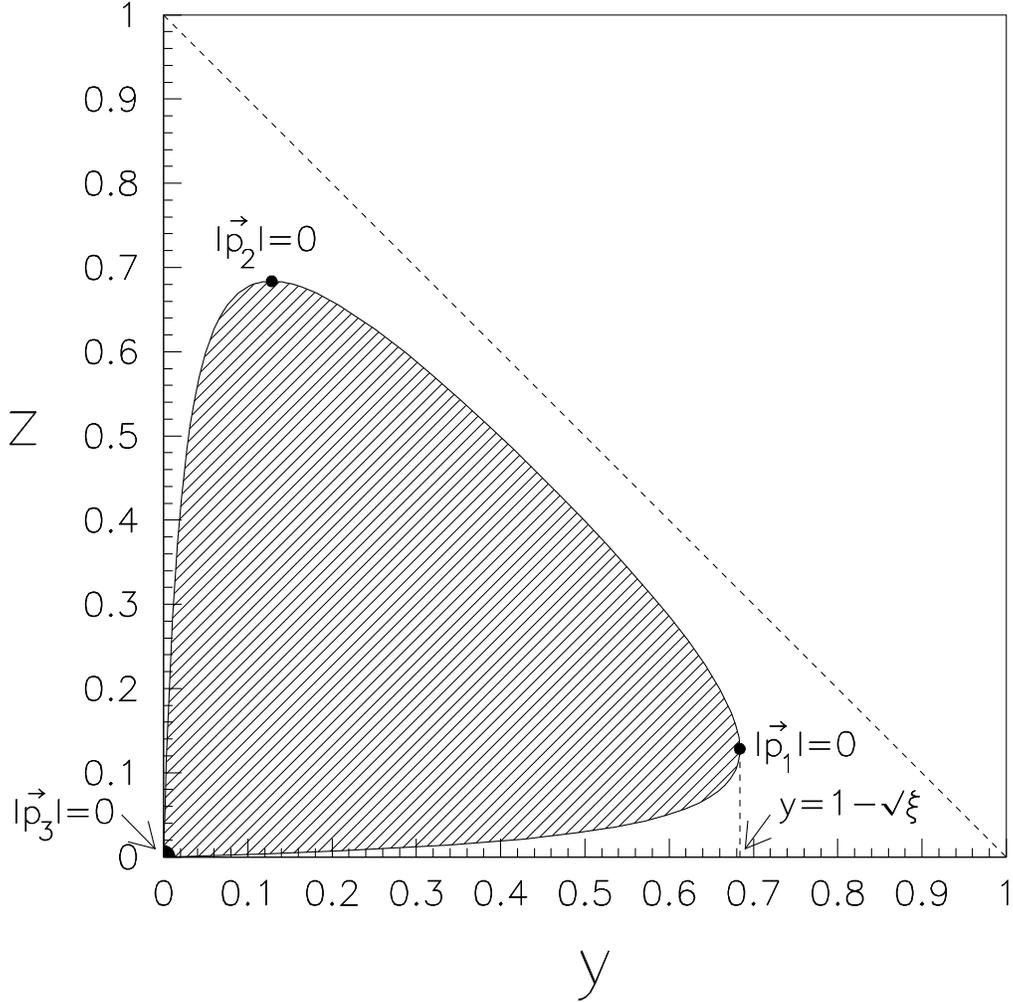, scale=0.8}
\caption{\label{fig4}Example for the phase space for the three particle decay}
\end{center}\end{figure}

\vspace{7pt}
The hadronic part of the diagrams is shown in Fig.~\ref{fig3}. The additional
degree of freedom given by the gluon leads to two additional integration
measures, expressible in the energy type variables $y$ and $z$ mentioned
earlier. The three particle phase space for some specified center-of-mass
energy $\sqrt{q^2}$ is shown in Fig.~\ref{fig4}.
Integrals to be calculated are of the general type ($p$ and $q$ take integer
values in a finite range around zero)
\begin{equation}
I(\xi,\Lambda)=\int y^pz^qdy\,dz
\end{equation}
where the integration limits are given by the phase space boundary,
\begin{eqnarray}
y_-&=&\sqrt{\Lambda\xi}+\Lambda,\qquad y_+\ =\ 1-\sqrt\xi\\
z_\pm(y)&=&\frac{2y}{4y+\xi}\left\{1-y-\frac12\xi+\Lambda+\frac\Lambda y
  \pm\frac1y\sqrt{(y-\Lambda)^2-\Lambda\xi}\sqrt{(1-y)^2-\xi}\right\}
\end{eqnarray}
where $\sqrt{\Lambda q^2}$ is again the gluon mass. While the integration
over $z$ can be performed easily, the second integration is not so easy to
perform, the most complicated (logarithmic) integrals are of the shape
\begin{equation}
I'(\xi,\Lambda)=\int_{\sqrt{\Lambda\xi}}^{1-\sqrt\xi}
  y^n\ln\pfrac{z_+(y)}{z_-(y)}dy.
\end{equation}
Two issues should be mentioned here. First, the gluon mass is necessary only
for regularizing the integrals. Therefore, the parameter $\Lambda$ need to
appear only at places where the integral at the lower limit would lead to
singular expressions if the parameter were absent. A lot of integrals,
therefore, can be calculated for $\Lambda=0$. On the other hand, the
singular integrals can be calculated by subtracting and adding a (simpler)
integral which has the same singular structure. According to
\begin{equation}
I'(\xi,\Lambda)=\left(I'(\xi,\Lambda)-I'_0(\xi,\Lambda)\right)
  +I'_0(\xi,\Lambda)=\Delta I'(\xi,\Lambda)+I'_0(\xi,\Lambda),
\end{equation}
for the difference term $\Lambda=0$ can be chosen. The integral $I'_0$ can be
obtained from $I'$ by taking approximations for the integrand close to the
lower (infrared singular) limit, $y=\sqrt{\Lambda\xi}+\Lambda$. So
\begin{equation}
z_\pm\rightarrow\frac1\xi\left((2-\xi)y\pm2\sqrt{(y-\Lambda)^2-\Lambda\xi}
  \sqrt{1-\xi}\right)
\end{equation}
and a substitution $y=\Lambda+\sqrt{\Lambda\xi}\cosh\zeta$ helps for the
integration. For the integrals with $\Lambda=0$, on the other hand, a combined
substitution $y=1-\sxi(1+z^2)/(1-z^2)$ works to separate the integral into
integrable pieces. The results can be finally expressed in a very compact form
by using the so-called decay rate terms. At this point I only show the results
for the unpolarized case for reasons of brevity, the spin-dependent parts are
only different but not more complicated, ($N=\alpha_sN_cC_f/4\pi v$)
\begin{eqnarray}
H_U^1(\alpha_s)&=&N\Big\{2(2+7\xi)v+(48-48\xi+7\xi^2)t_3
  +2\sxi\left(2-7\xi+\sxi(2+3\xi)\right)t_4\nonumber\\&&
  -2\xi(2+3\xi)t_5-4(2-\xi)\left((2-\xi)(t_8-t_9)+2v(t_{10}+2t_{12})\right)
  \Big\},\nonumber\\
H_U^2(\alpha_s)&=&\xi N\Big\{12v+2(6-\xi)t_3+2\sxi(1-\sxi)t_4\nonumber\\&&
  +2\xi t_5-4\left((2-\xi)(t_8-t_9)+2v(t_{10}+2t_{12})\right)\Big\},\nonumber\\
H_L^1(\alpha_s)&=&\!\!N\Big\{\frac12(16-46\xi+3\xi^2)v
  +\frac\xi4(88-32\xi+3\xi^2)t_3\nonumber\\&&
  -2\sxi\left(2-7\xi+\sxi(2+3\xi)\right)t_4+2\xi(2+3\xi)t_5\nonumber\\&&
  -2\xi\left((2-\xi)(t_8-t_9)+2v(t_{10}+2t_{12})\right)\Big\},\nonumber\\
H_L^2(\alpha_s)&=&\xi N\Big\{\frac32(10-\xi)v
  +\frac14(24-16\xi-3\xi^2)t_3\nonumber\\&&
  -2\sxi(1-\sxi)t_4-2\xi t_5
  -2\left((2-\xi)(t_8-t_9)+2v(t_{10}+2t_{12})\right)\Big\},\nonumber\\[3pt]
H_F^3(\alpha_s)&=&-8\xi N v\pi,\nonumber\\[3pt]
H_F^4(\alpha_s)&=&N\Big\{-16\sxi(1-\sxi)-16(t_1-t_2)+8(2-3\xi)vt_3\nonumber\\&&
  -4(4-5\xi)t_6-8v\left((2-\xi)(t_8-t_7)+2v(t_{10}+t_{11})\right)\Big\}
\end{eqnarray}
where the decay rate terms, containing logarithms and dilogarithms, are found
in Ref.~\cite{GKT97} together with the polarized contributions. More
interesting at this point are the results we obtain. In Fig.~\ref{fig5} (top)
the total cross section for pair production of top quarks is shown in
dependence on the center-of-mass energy $\sqrt{q^2}$, with and without the
first order radiative correction. The polar angle dependence of the
longitudinal polarization
\begin{equation}
P=\frac{d\sigma(\uparrow)-d\sigma(\downarrow)}{d\sigma(\uparrow)
  +d\sigma(\downarrow)}
\end{equation}
of the top quark is shown in Fig.~\ref{fig5} (bottom). In Fig.~\ref{fig6} the
polar angle dependence of the normal (top) and perpendicular polarization
(bottom) of the top quark is displayed. What is not shown here is the change
with respect to the Born term contribution. It figures out that despite the
fact of rather large corrections to the total cross section, the corrections
to the polarizations are quite small (about 5\%). A detailed discussion of the
results can be found in Refs.~\cite{GKT96,GK96,GKT97}. Quantities like the
polarization can be measured by the asymmetry of the decay products of the
top quark.

\subsection{Finite gluon energy cut}
Besides the comparison with the massless case, resulting in anomalous
spin-flip contributions which will be shown later, additional work has been
investigated in the exact calculation of the contributions for a given gluon
energy cut, dividing the phasespace up into a soft and a hard
part~\cite{GKK98}. This ``excurse'' will not be taken here.

\section{Longitudinal spin correlation}
If both quark and antiquark are considered to be polarized, the correlation
between their spins can be measured. In Ref.~\cite{GKL98} the correlation of
longitudinal polarized quark and antiquark is considered, in Ref.~\cite{GKL00}
the angular dependence is considered as well. Because of the representation
\begin{eqnarray}
(s_1^\ell)^\mu&=&\frac{s_1^\ell}{\sqrt\xi}(\sqrt{(1-y)^2-\xi};0,0,1-y)\\
(s_2^\ell)^\mu&=&\frac{s_2^\ell}{\sqrt\xi}(\sqrt{(1-z)^2-\xi};
  (1-z)\sin\theta_{12},0,(1-z)\cos\theta_{12})\nonumber
\end{eqnarray}
for the longitudinal spins where
\begin{equation}
\cos\theta_{12}=\frac{yz+y+z-1+\xi}{\sqrt{(1-y)^2-\xi}\sqrt{(1-z)^2-\xi}},
\end{equation}
there occurs an additional square root $\sqrt{(1-z)^2-\xi}$ in the
denominator which makes the integrations more complicated. However, the same
integration methods work also in this case, an appropiate substitution could
be found to make the integration feasible. New decay rate terms had to be
defined in order to describe the final result (see Ref.~\cite{GKL00}). 

\begin{figure}[ht]\begin{center}
\epsfig{figure=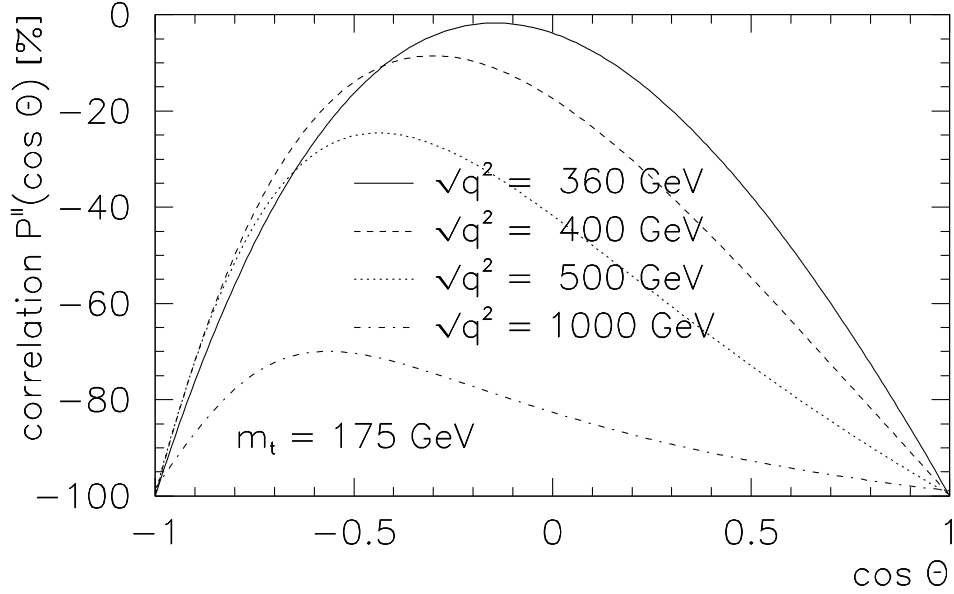, scale=0.8}
\caption{\label{fig7}Correlation of longitudinal top and antitop spin}
\end{center}\end{figure}

\subsection{Result for the correlation}
The correlation is defined as
\begin{equation}
C=\frac{d\sigma(\uparrow\uparrow)-d\sigma(\uparrow\downarrow)
  -d\sigma(\downarrow\uparrow)+d\sigma(\downarrow\downarrow)}
  {d\sigma(\uparrow\uparrow)+d\sigma(\uparrow\downarrow)
  +d\sigma(\downarrow\uparrow)+d\sigma(\downarrow\downarrow)}
\end{equation}
The polar angle dependence of this result is shown in Fig.~\ref{fig7}.

\subsection{Anomalous spin-flip contributions}
It is an easy task to calculate results for polarized structure functions
also in the case where the quark mass is assumed to be massless, even though
this calculation have to be performed in dimensional regularization. The
results, however, differ from results one obtains by taking the results for
the massive quark as obtained earlier and calculating the limit $m\to 0$.
By adding in the Born term contributions one finally obtains in the $m\to 0$
limit ($C_F=4/3$ is made explicit here)
\begin{eqnarray}\label{anomalies}
H^1_U(s_1^\ell,s_2^\ell)
  &=&\frac14\Big(H^1_U+H^{1(\ell_1\ell_2)}_Us_1^\ell s_2^\ell\Big)
  \ =\nonumber\\
  &=&N_cq^2\left((1-s_1^\ell s_2^\ell)
  \left(1+\frac13\times\frac{\alpha_s}\pi\right)
  +\left[\frac43\times\frac{\alpha_s}\pi s_1^\ell s_2^\ell\right]\right),
  \nonumber\\
H^1_L(s_1^\ell,s_2^\ell)
  &=&\frac14\Big(H^1_L+H^{1(\ell_1\ell_2)}_Ls_1^\ell s_2^\ell\Big)
  \ =\nonumber\\
  &=&N_cq^2(1-s_1^\ell s_2^\ell)\left(0+\frac23\times\frac{\alpha_s}\pi
  +[0]\right),\nonumber\\
H^1_F(s_1^\ell,s_2^\ell)
  &=&\frac14\Big(H^{1(\ell_1)}_Fs_1^\ell+H^{1(\ell_2)}_Fs_2^\ell\Big)
  \ =\nonumber\\
  &=&N_cq^2(s_1^\ell-s_2^\ell)\left(1+0\times\frac{\alpha_s}\pi
  -\left[\frac23\times\frac{\alpha_s}\pi\right]\right),\nonumber\\ 
H^4_U(s_1^\ell,s_2^\ell)
  &=&\frac14\Big(H^{1(\ell_1)}_Us_1^\ell+H^{1(\ell_2)}_Us_2^\ell\Big)
  \ =\nonumber\\
  &=&N_cq^2(s_1^\ell-s_2^\ell)\left(1+\frac13\times\frac{\alpha_s}\pi
  -\left[\frac23\times\frac{\alpha_s}\pi\right]\right),\nonumber\\ 
H^4_L(s_1^\ell,s_2^\ell)
  &=&\frac14\Big(H^{1(\ell_1)}_Ls_1^\ell+H^{1(\ell_2)}_Ls_2^\ell\Big)
  \ =\nonumber\\
  &=&N_cq^2(s_1^\ell-s_2^\ell)\left(0+\frac23\times\frac{\alpha_s}\pi
  +[0]\right),\nonumber\\
H^4_F(s_1^\ell,s_2^\ell)
  &=&\frac14\Big(H^4_F+H^{4(\ell_1\ell_2)}_Fs_1^\ell s_2^\ell\Big)
  \ =\nonumber\\
  &=&N_cq^2\left((1-s_1^\ell s_2^\ell)
  \left(1+0\times\frac{\alpha_s}\pi\right)
  +\left[\frac43\times\frac{\alpha_s}\pi s_1^\ell s_2^\ell\right]\right)
\end{eqnarray}
where the anomalous spin-flip contributions are made explicit by the square
bracket notation. These contributions have their origin in the collinear limit
where the spin-flip contribution proportional to $m$ survives since it is
multiplied by the $1/m$ collinear mass singularity. Because the anomalous
spin-flip terms are associated with the collinear singularity, the flip
contributions are universal and factorize into the Born term contribution and
an universal spin-flip bremsstrahlung function~\cite{FalkSehgal}. This
explains why there is no anomalous contribution to $H_L^4$ and why the
anomalous flip contributions to $H_U^4$ and $H_F^1$, for instance, are equal.
In fact the strength of the anomalous spin-flip contribution can directly be
calculated from the universal helicity-flip bremsstrahlung function listed in
Ref.~\cite{FalkSehgal}. 

\begin{figure}[ht]\begin{center}
\epsfig{figure=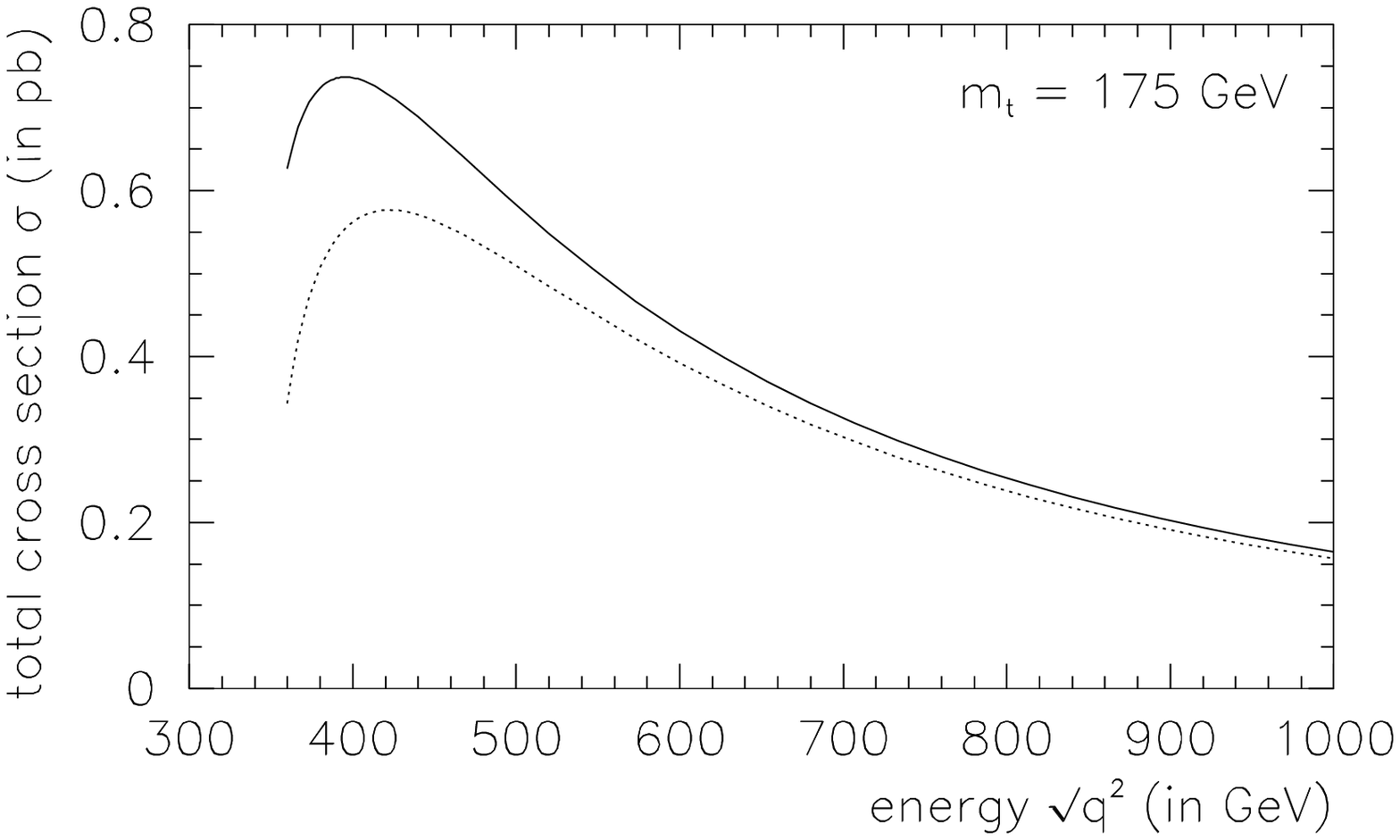, scale=0.8}\qquad
\epsfig{figure=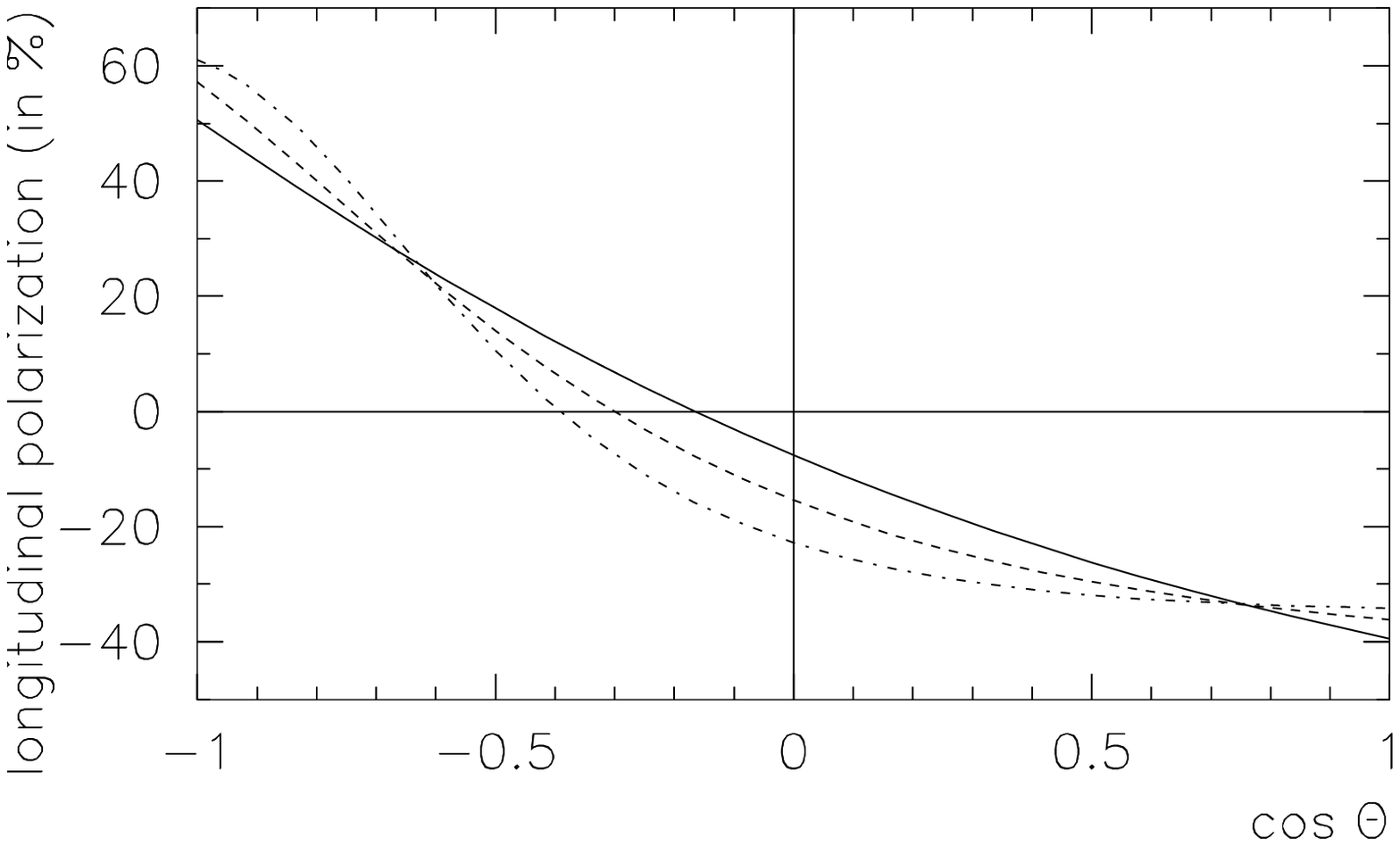, scale=0.8}
\caption{\label{fig5}total cross section dependence on the center-of-mass
  energy $\sqrt{q^2}$ with and without first order radiative corrections for
  the top quark pair production (top) and polar angle dependence of the
  longitudinal polarization for the center-of-mass energies $380\GeV$ (solid),
  $500\GeV$ (dashed) and $1000\GeV$ (dashed-dotted) including radiative
  corrections}
\end{center}\end{figure}
\begin{figure}[ht]\begin{center}
\epsfig{figure=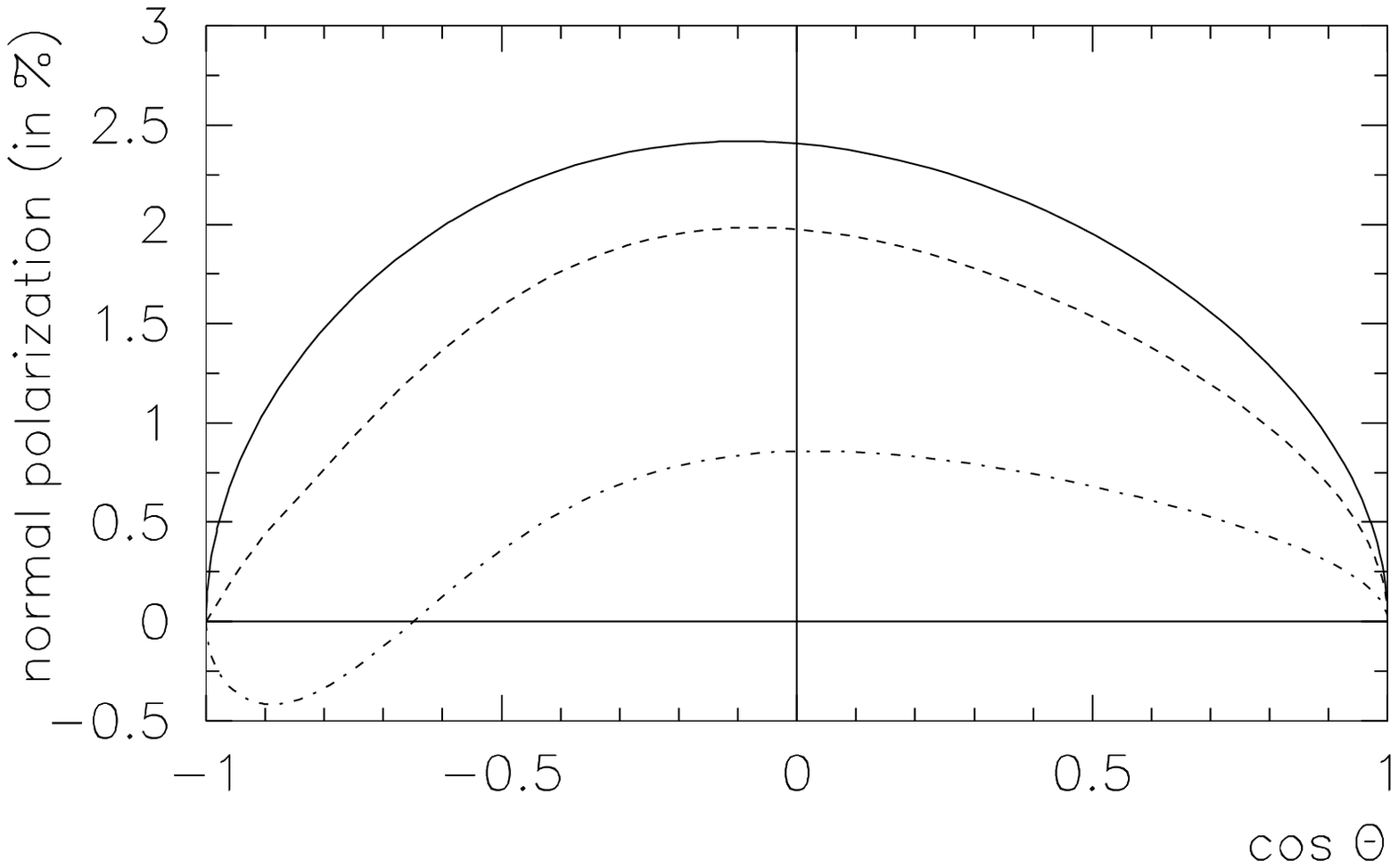, scale=0.8}\qquad
\epsfig{figure=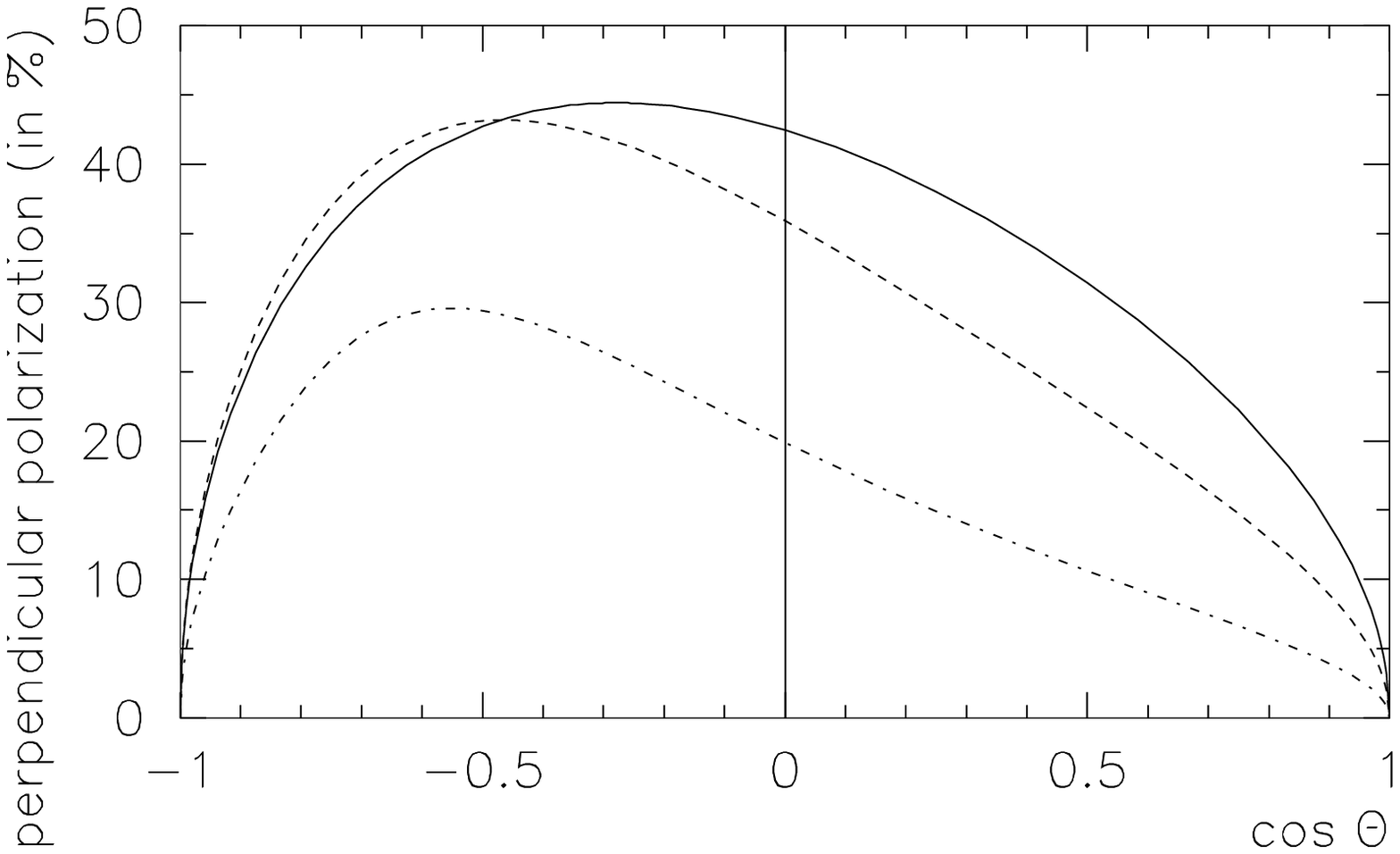, scale=0.8}
\caption{\label{fig6}angular dependence of the normal polarization (top) and
  the perpendicular polarization of the top quark (bottom) for the
  center-of-mass energies $380\GeV$ (solid), $500\GeV$ (dashed) and $1000\GeV$
  (dashed-dotted) including first order radiative corrections}
\end{center}\end{figure}

\section{Gluon polarization}
The polarization of gluons in $e^+e^-$-annihilation~\cite{OlsenOsland,KS}, in
deep inelastic scattering~\cite{OlsenOlsen} and in quarkonium
decays~\cite{OlsenOsland,Brodsky} has been studied in a series of papers
dating back to the early 80's. Several proposals have been made to measure the
polarization of the gluon among which is the analytic proposal to angular
correlation effects in the splitting process of a polarized gluon into a
pair of gluons or quarks.

\vspace{7pt}
Results for the linear and circular polarization of gluons produced in the
$e^+e^-$ annihilation process for massive quarks with subsequent gluon
emission are presented in this section. Similar to the joined quark-antiquark
density matrix the two-by-two differential density matrix
$d\ssigma=d\sigma_{\lambda_G\lambda'_G}$ of the gluon is used with gluon
helicities $\lambda_G=\pm 1$ in terms of its components along the unit matrix
and the three Pauli matrices. Accordingly one has
\begin{equation}\label{eqn1}
d\ssigma=\frac12(d\sigma\oone+d\sigma^x\ssigma_x+d\sigma^y\ssigma_y
  +d\sigma^z\ssigma_z),
\end{equation}
where $d\sigma$ is the unpolarized differential rate and 
$d\vec\sigma=(d\sigma^x,d\sigma^y,d\sigma^z)$ are the three components of 
the (unnormalized) differential Stokes vector. After azimuthal averaging the
$y$-component of the Stokes vektor $d\sigma^y$ drops out, one is left with the
$x$- and 
$z$-components of the Stokes vector which are referred to as the gluon's 
linear polarization in the event plane and the circular polarization of 
the gluon, respectively.

\subsection{The differential cross section}
The differential unpolarized and polarized cross section, differential with
regard to the polar beam-event orientation and the two energy-type variables
$x=2p_3\cdot q/q^2$ and $w=2(p_1-p_2)\cdot q/q^2$ (with $q=p_1+p_2+p_3$) are
then given by
\begin{eqnarray}
\frac{d\sigma^{(x)}}{d\cos\theta\,dx\,dw}
  &=&\frac38(1+\cos^2\theta)\left(g_{11}\frac{d\sigma_U^{1(x)}}{dx\,dw}
  +g_{12}\frac{d\sigma_U^{2(x)}}{dx\,dw}\right)\label{eqn2}\,+\\&&
  +\frac34\sin^2\theta\left(g_{11}\frac{d\sigma_L^{1(x)}}{dx\,dw}
  +g_{12}\frac{d\sigma_L^{2(x)}}{dx\,dw}\right)
  +\frac34\cos\theta\,g_{44}\frac{d\sigma_F^{4(x)}}{dx\,dw},\nonumber\\
\frac{d\sigma^z}{d\cos\theta\,dx\,dw}
  &=&\frac38(1+\cos^2\theta)g_{14}\frac{d\sigma_U^{4z}}{dx\,dw}
  +\frac34\cos\theta\left(g_{41}\frac{d\sigma_F^{1z}}{dx\,dw}
  +g_{42}\frac{d\sigma_F^{2z}}{dx\,dw}\right).\qquad\label{eqn3}
\end{eqnarray}
The notation $d\sigma^{(x)}$ stand for either $d\sigma$ or $d\sigma^x$, and 
the same for $d\sigma_\alpha^{i(x)}$. Of course, in this case no Born terms
occur, the first order tree diagram is the leading order contribution. The
phase space in terms of the variables $x$ and $w$ is similar to the one in
$y$ and $z$ (see Fig.~\ref{fig4}), standing on the sharp corner. The phase
space limits are therefore symmetric,
\begin{eqnarray}
x_-&=&0,\qquad
x_+\ =\ 1-\xi,\nonumber\\
w_\pm(x)&=&\pm x\sqrt{\frac{1-2x-\xi}{1-2x}}.
\end{eqnarray}

\subsection{Results for the gluon polarization}
\begin{figure}[ht]\begin{center} 
\epsfig{figure=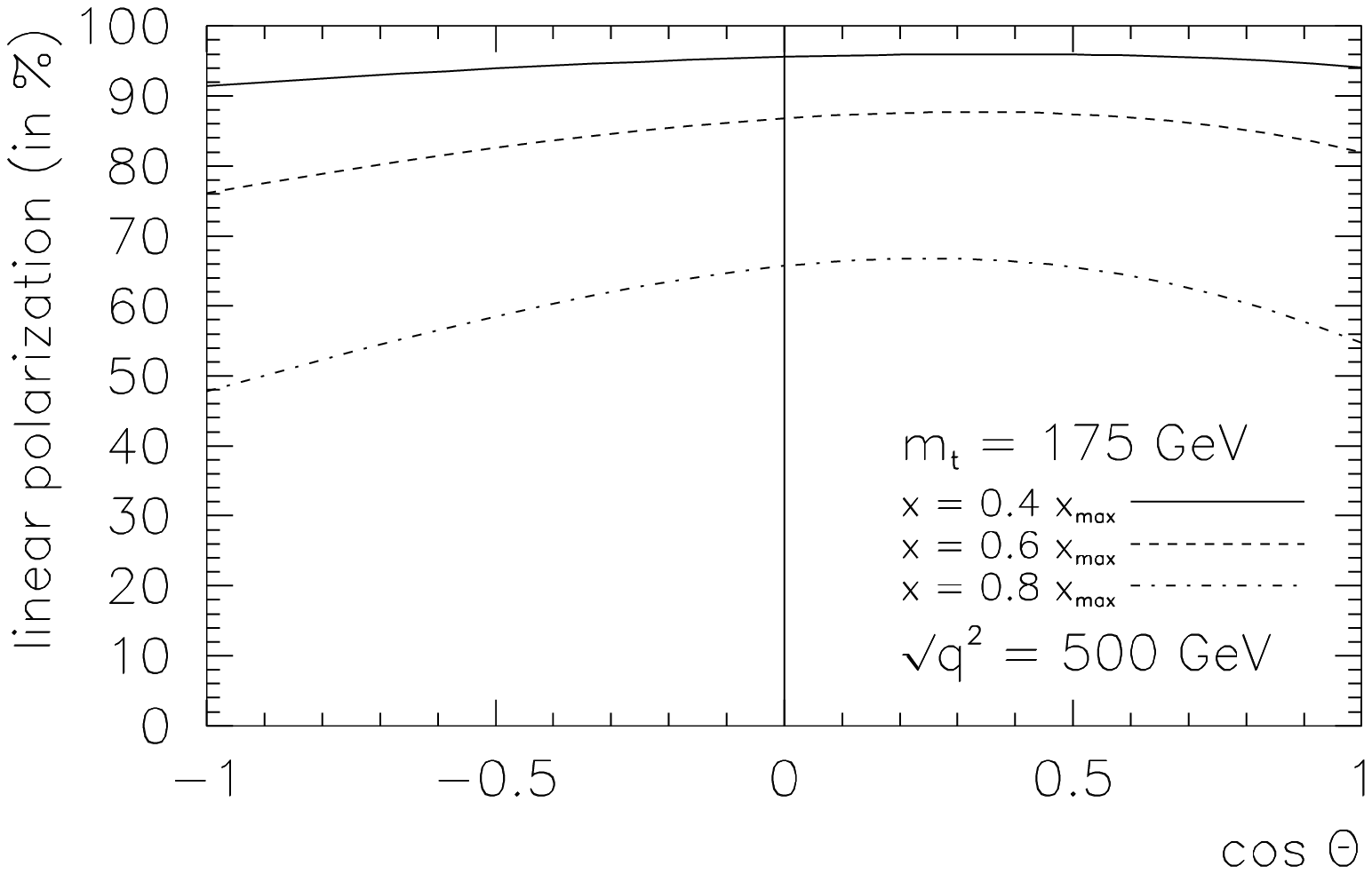, scale=0.8}
\epsfig{figure=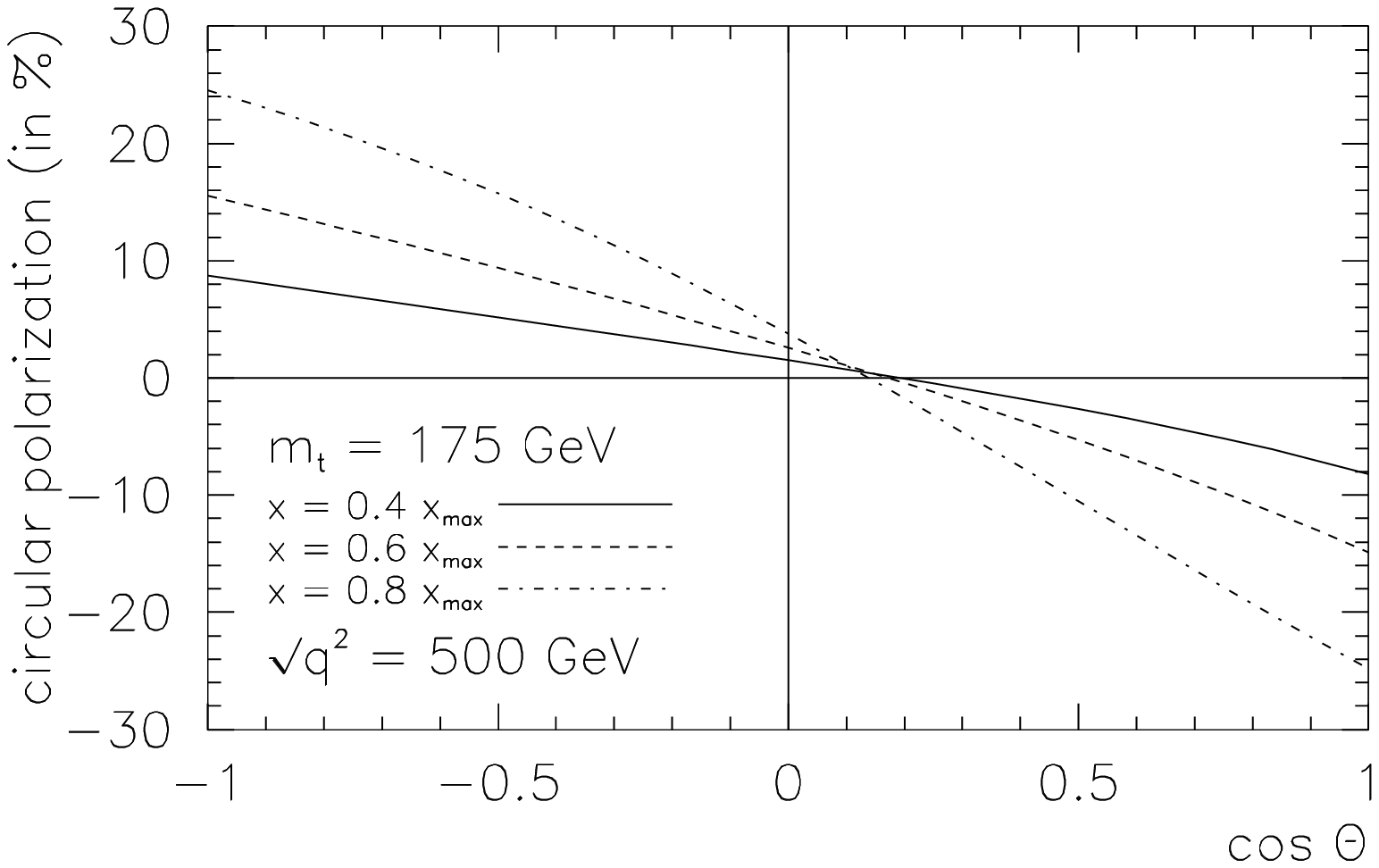, scale=0.8}
\caption{\label{fig8}polar angle dependence of the linear (top) and circular
  polarization (bottom) of the gluon in the case of top quark pair production}
\end{center}\end{figure}

The integration can be done also in this case by using an appropiate
substitution (see Refs.~\cite{GKL97,GKL99}). However, it is also worth to
consider the results still depending on the gluon energy variable $x$. The
results are given by
\begin{eqnarray}
H_U^1(x)&=&-32\Big(\frac4x-4+x\Big)\frac1xw_+(x)
  +32\Big(\frac2x-2+x\Big)t_+^\ell(x),\nonumber\\
H_U^2(x)&=&-32\xi\Big(\frac1x-1\Big)\frac1xw_+(x)
  +16\xi\Big(\frac{2-\xi}x-2\Big)t_+^\ell(x),\nonumber\\
H_L^1(x)&=&16(4+\xi)\Big(\frac1x-1\Big)\frac1xw_+(x)
  -8\xi\Big(\frac{6-\xi}x-2-x\Big)t_+^\ell(x),\\
H_L^2(x)&=&-16\xi\Big(\frac1x-1\Big)\frac1xw_+(x)
  +8\xi\Big(\frac{2-\xi}x-2-x\Big)t_+^\ell(x),\nonumber\\
H_F^4(x)&=&64\Big(\frac{1-\xi}x-1\Big)
  -32\Big(\frac2x-2+x\Big)t_-^\ell(x),\nonumber\\[12pt]
H_U^{1x}(x)&=&-128\Big(\frac1x-1\Big)\frac1xw_+(x)
  +64\Big(\frac1x-1)t_+^\ell(x)\nonumber\\
H_U^{2x}(x)&=&-32\xi\Big(\frac1x-1\Big)\frac1xw_+(x)
  +16\xi\Big(\frac{2-\xi}x-2\Big)t_+^\ell(x),\nonumber\\
H_L^{1x}(x)&=&16(4+\xi)\Big(\frac1x-1\Big)\frac1xw_+(x)
  -8\xi\Big(\frac{6-\xi}x-2\Big)t_+^\ell(x),\\
H_L^{2x}(x)&=&-16\xi\Big(\frac1x-1\Big)\frac1xw_+(x)
  +8\xi\Big(\frac{2-\xi}x-2\Big)t_+^\ell(x),\nonumber\\
H_F^{4x}(x)&=&64\Big(\frac{1-\xi}x-1\Big)
  -64\Big(\frac1x-1\Big)t_-^\ell(x),\nonumber\\[12pt]
H_U^{4z}(x)&=&64(1-\xi-x)-32(2-x)t_-^\ell(x),\qquad
H_L^{4z}(x)\ =\ 0,\nonumber\\[3pt]
H_F^{1z}(x)&=&-32(4-3x)\frac1xw_+(x)+32(2-x)t_+^\ell(x),\qquad
H_F^{2z}(x)\ =\ 0
\end{eqnarray}
where
\begin{equation}
w_+(x)=x\sqrt{\frac{1-x-\xi}{1-x}},\quad
t_+^\ell(x)=\ln\left(\frac{\sqrt{1-x}
  +\sqrt{1-x-\xi}}{\sqrt{1-x}-\sqrt{1-x-\xi}}\right),\quad
t_-^\ell(x)=\ln\pfrac{1-x}\xi.
\end{equation}
For different values of $x$ in terms of $x_{\rm max}=1-\xi$ the polar angle
dependence of the linear and circular polarization of the top quark is shown
in Fig.~\ref{fig8}.

\section{Conclusion}
First order radiative QCD corrections for polarization observables can be done
analytically and are done in most of the cases. The corrections for the
polarization observables figure out to be small. This, however, proves the
quality of perturbation theory estimates. The polarization observables can be
measured by the asymmetry in the decay rate of final state leptons. The
consideration of massive quarks are important especially at the high energies
which are reached by the NLC. The calculation techniques can be combined with
other techniques like the high spin tools developed in the theory group at
Tartu University, a future collaboration is scheduled.

\subsection*{Acknowledgements:}
I want to thank my J.G.~K\"orner, M.M.~Tung, J.A.~Leyva, and V.~Kleinschmidt
for a fruitful collaboration on this field. My work is supported by a
habilitation grant given by the DFG.

\end{document}